\documentclass[reprint,amsmath,amssymb,aps,abbrv,superscriptaddress]{revtex4-1}
\usepackage{cancel}
\usepackage{mciteplus,physics}
\usepackage{graphicx,cancel,float}
\usepackage{dcolumn}
\usepackage[caption=false]{subfig}
\usepackage{bm}

\usepackage{braket,comment}
\graphicspath{{Figures/}}
\usepackage{hyperref}
\usepackage[normalem]{ulem}
\usepackage[dvipsnames]{xcolor}
\usepackage[capitalise]{cleveref}

 \newcommand{\sgn}{\textrm{sgn} }
 \newcommand{\tchi}{\chi}
\newcommand{\tS}{S}
  \newcommand{\tv}{v}
\newcommand{\tu}{u}
\newcommand{\tTil}{\tau}
\newcommand{\tTau}{\mathcal{T}}
\newcommand{\tH}{\widetilde{H}}
\newcommand{\trho}{\rho}
\newcommand{\tPsi}{\Psi}
\def\iu{\rm{i}}
\def\e{\rm{e}}
\newcommand{\citer}[1]{Ref. \cite{#1}}

\begin{document}
\title{Violent relaxation in quantum fluids with long-range interactions}
\date{\today}
\author{Ryan Plestid}
\email{plestird@mcmaster.ca}
\affiliation{Department of Physics and Astronomy, McMaster University, 
1280 Main St.\ W.\, Hamilton, Ontario, Canada}
\affiliation{Perimeter Institute for Theoretical Physics, 31 Caroline St. N.,
 Waterloo, Ontario, Canada}
\author{Perry Mahon}
\affiliation{Department of Physics and Astronomy, McMaster University, 
1280 Main St.\ W.\, Hamilton, Ontario, Canada}
\affiliation{Department of Physics, University of Toronto, 
  60 St.\ George St., Toronto, Ontario, Canada}
\author{D.H.J. O'Dell}
\email{dodell@mcmaster.ca}
\affiliation{Department of Physics and Astronomy, McMaster University, 
1280 Main St.\ W.\, Hamilton, Ontario, Canada}

\begin{abstract}
Violent relaxation is a process that occurs in systems with long-range interactions. It has the peculiar feature of dramatically amplifying small perturbations, and rather than driving the system to equilibrium it instead leads to slowly evolving configurations known as quasi-stationary states that fall outside the standard paradigm of statistical mechanics. Violent relaxation was originally identified in gravity-driven stellar dynamics; here we extend the theory into the quantum regime by developing a quantum version of the Hamiltonian Mean Field (HMF) model which exemplifies many of the generic properties of long-range interacting systems. The HMF model can either be viewed as describing particles interacting via a cosine potential, or equivalently as the kinetic XY-model with infinite range interactions, and its quantum fluid dynamics can be obtained from a generalized Gross-Pitaevskii equation. We show that singular caustics that form during violent relaxation are regulated by interference effects in a universal way described by Thom's catastrophe theory applied to waves and this leads to emergent length and time scales not present in the classical problem. In the deep quantum regime we find that violent relaxation is suppressed altogether by quantum zero-point motion. Our results are relevant to laboratory studies of self-organization in cold atomic gases with long-range interactions.
\end{abstract}

\maketitle


\section{Introduction\label{sec:Intro}}
Quantum many-body (QMB) systems with long-range interactions (LRI) are increasingly being realized in laboratory experiments with cold atomic and molecular gases where inherently long coherence times are suited to investigating dynamics. Examples include atomic Bose-Einstein condensates (BEC) with magnetic dipole-dipole interactions \cite{Griesmaier05,Lahaye07,Lahaye08,Koch08,Beaufils08,Lu11,Pasquiou11,Aikawa12,Kadau16,Ferrier-Barbut2016, Chomaz2017,Wenzel2017}, cold polar molecules \cite{Chotia12,Yan13,Martin13,Hazzard14,Richerme2014,Moses15}, trapped ions \cite{Kim09,Britton12,Korenbilt12,Islam13}, Rydberg atoms \cite{Heidemann09,Comparat10,Schwarzkopf11,Schauss12,Gunter13,Schempp15,Barredo15,Labuhn16}, and atoms inside high-finesse optical cavities which interact via the cavity modes that extend over the entire cavity \cite{Black03,Baumann10,Mottl12,Landig15,Landig2016,Leonard17}. There are also new approaches in the pipeline, such as using optical waveguides or photonic bandgap crystals to engineer electromagnetic modes and hence mediate highly controlled long-range interatomic interactions  \cite{Douglas15,Gonzalez15,Shahmoon2016}. 

This progress in trapped atomic and molecular systems has fostered broad interest in LRI both in and out of equilibrium \cite{Porras04,Gopalakrishnan11,Hazzard13,Hauke13,Eisert2013,Shachenmayer2013,Grass2014,vanBijnen15,Maghrebi2017,Zeiher16,Glaetzle17}. While the focus has been on spin and Hubbard models, the versatility of these systems also allows for new regimes not seen in traditional condensed matter systems, including gravity-like attractive $1/r$ interactions \cite{Odell00,Giovanazzi01,Kurizki02}, and also cosine-type interactions that occur between atoms in optical cavities \cite{Domokos01,Domokos02,Asboth05,Nagy10,Keeling10,Ritsch13,Schutz2014,Schutz2016,Keller2017}.

Historically, the motivation for studying LRI has come from astrophysics and plasma physics: The range of the gravitational and Coulomb interactions, respectively, are such that all particles experience a common, essentially mean-field, potential. In non-equilibrium situations this potential becomes time dependent and drives a rapid, collisionless relaxation mechanism, known as violent relaxation, which efficiently mixes phase-space \cite{Binney2011}. This process is non-ergodic and hence profoundly different from relaxation in systems with short-range interactions. Nevertheless, universality still emerges: pioneering work in the 1960s by Lynden-Bell \cite{Lynden-Bell1967}  on the statistical mechanics of violent relaxation in stellar and galactic dynamics introduced a fourth type of equilibrium distribution which is related to both the Fermi-Dirac distribution and equipartition of energy per unit mass.  More recent research has generalized  Lynden-Bell statistics to two-parameter Core-Halo distributions \cite{Yamaguchi2008,Levin2008,Pakter2011,Teles2013} which can also handle the case of far-from-virialized initial conditions and the following two-stage picture of relaxation from a non-equilibrium state has emerged \cite{Campa2009,Levin2014}: Firstly there is violent relaxation, the timescale of which does not depend on the number of particles $N$, and  results in \emph{long-lived non-equilibrium configurations} known as quasi-stationary states (QSS). Secondly, at long times, there is the more familiar collisional relaxation towards Maxwell-Boltzmann equilibrium, however this occurs at times of order $N^{\delta}$, where $\delta\ge 1$   \cite{Yamaguchi2004,Jain07}. Therefore, in the thermodynamic limit $N \rightarrow \infty$ the lifetime of the QSS diverges and the system remains out of equilibrium indefinitely which has implications for thermalization. 
Violent relaxation is now recognized as the cornerstone of statistical theories describing the QSS that dominate transient behaviour in systems with LRI. Since these QSS are formed by non-ergodic dynamics they are not captured by conventional statistical treatments.


\begin{figure}[!h]
\includegraphics[width=\linewidth]{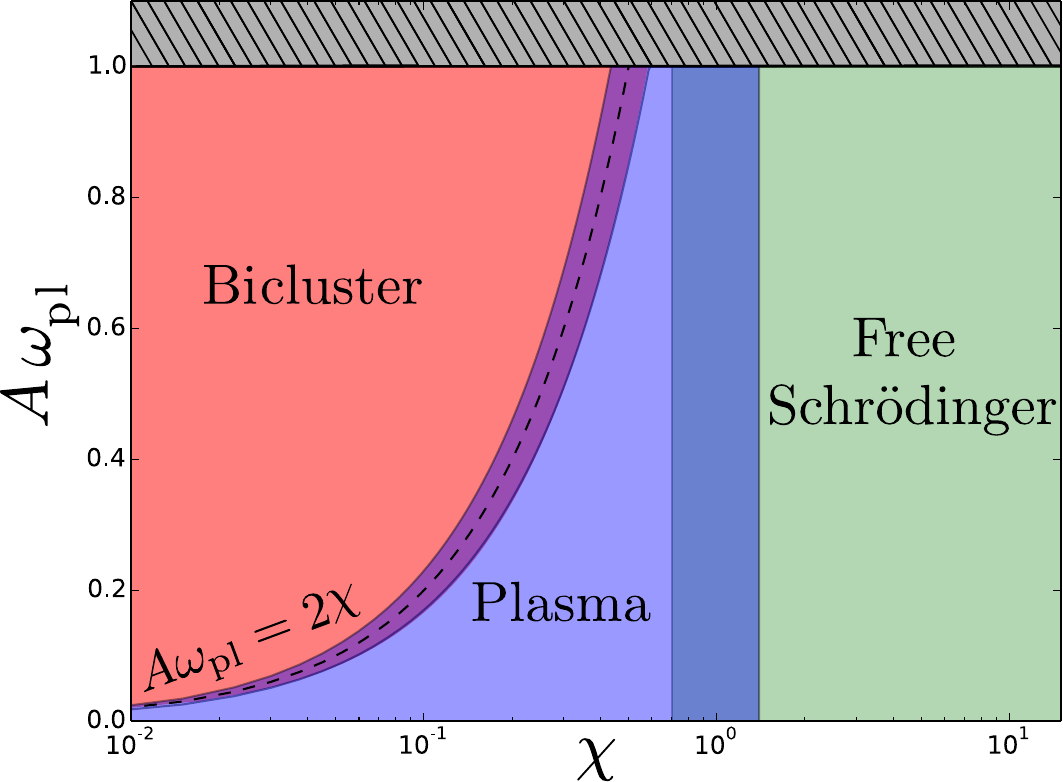}
\caption{Non-equilibrium phase diagram for the quantum HMF model with repulsive interactions ($\epsilon >0$). Violent relaxation leads to quasi-stationary states which are very slowly evolving non-equilibrium configurations; for the HMF model these are the bicluster states. On this semi-log plot, the vertical axis measures the initial deviation from equilibrium quantified by $\tv_0=A\omega_\text{pl} \cos\theta$ which is the initial velocity field due to small plasma oscillations [see \cref{eq:linsolns}]. The horizontal axis captures quantum effects via the effective Planck constant  $\tchi=\hbar/\sqrt{\vert \epsilon \vert m R^2}$. When $\tchi\ll 1$ \emph{and} $\tchi\lesssim 2 A\omega_\text{pl}$ the system forms a bicluster at late times (see \cref{fig:bi-d}). By contrast, for $\tchi\lesssim 1$, the $k=\pm 1$ modes dominate and the system continues to undergo plasma oscillations without relaxing (see \cref{eq3:Lin-Qu} and \cref{fig:plas}). Finally, for $\tchi\gtrsim 1$ all Fourier modes experience a free Schr\"{o}dinger-like dispersion relation and violent relaxation is suppressed by quantum zero-point motion (see \cref{fig:schro}). The grey region denotes initial conditions that invalidate the short-time linear response procedure detailed in \cref{eq:cl-multi-scale,eq:qu-multi-scale}. \label{fig:phase-diagram}}
\end{figure}

In this paper we are interested in the following thematic questions: Does violent relaxation take place in quantum systems? Do QSS exist and do they display new length and time scales? Do these modifications  survive in the thermodynamic limit?  We base our analysis on the Hamiltonian Mean Field (HMF) model \cite{Antoni1995, Dauxois2002, Barre2002, Barre2002a, Jain07,Staniscia2011} which over the last two decades has become one of the main theoretical tools for investigating many-body systems with LRI---see References \cite{Campa2009} and \cite{Levin2014} for reviews. It offers the advantage of being analytically tractable at equilibrium, and is known to capture dynamical features  present in more complicated systems \cite{Dauxois2002,Barre2002a,Pakter2011,Levin2014}. Moreover, the HMF model is directly relevant to describing cold atoms in optical cavities where self-organization and the non-equilibrium Dicke phase transition have been intensively studied both theoretically \cite{Domokos01,Domokos02,Asboth05,Nagy10,Keeling10,Ritsch13,Schutz2014,Schutz2016,Keller2017}
and experimentally  \cite{Brennecke07,Colombe07,Slama07,Murch08,Baumann10,Wolke12}. Our work, therefore, has experimental relevance but for brevity's sake we only consider closed systems and do not include effects that would model cavity pumping and decay. However, in separate work we have shown that the type of QSS we observe (wave catastrophes) have the fundamental property of structural stability, even against decoherence, and hence survive in cavities weakly coupled to the environment \cite{Goldberg2016}. Wave catastrophes can also be seen in simulations by others \cite{Dominici2015} of microcavity polaritons using a driven damped Gross-Pitaevskii equation.

A key part of our results is summarized in the non-equilibrium phase diagram in Figure \ref{fig:phase-diagram} which depicts the end results of violent relaxation in a quantum version of the HMF model. The vertical axis gives the magnitude of initial perturbations from equilibrium, i.e.\ initial velocity fluctuations $\tv_0$, and the horizontal axis measures the effective Planck constant $\tchi$ which, of course, is entirely absent in classical systems.   We find that quantum effects increasingly stabilize initial plasma fluctuations, thereby suppressing violent relaxation, such that by the time the deep quantum regime (designated the free Schr\"{o}dinger phase) is reached quantum zero-point motion of higher lying modes dominates the plasma oscillations.

A second strand to the story we present here concerns the nature of the QSS and their connection to \emph{caustics}.  In a landmark paper on the large scale structure of the universe, Arnold, Shandarin and Zeldovich \cite{Arnold1982} considered a self-gravitating mass distribution and showed that initially smooth perturbations evolve into singular caustics upon which the density diverges. These caustics take universal shapes described by the so-called catastrophe theory due originally to Thom and Arnold \cite{Thom75,Arnold1975,Poston1978} and include cusps, swallowtails, beak-to-beaks, and Zeldovich pancakes \cite{Zeldovich70}. Likewise, as shown in \cref{fig:newtonian}, the dynamics of the classical HMF model also leads to cusp catastrophes \cite{Dauxois00,Firpo02,Barre2002,Barre2002a} which are the structurally stable catastrophes found in two dimensions (1 space + 1 time). 

One might expect quantum effects to smooth classical singularities and this is indeed what we find; however what is remarkable is that quantum effects enter in a universal way, replacing the caustics with characteristic interference patterns known as wave catastrophes  that introduce new length and time scales \cite{Berry1976,Berry1980}. For instance, cusp catastrophes become Pearcey functions (see Figure \ref{fig:pearcey}). Wave catastrophes obey a set of scaling relations as the wavelength is varied, and thus, as quantum effects are switched on \emph{the new length and time scales of the QSS scale in a universal way}. Once again, the non-ergodic nature of the classical limit plays a crucial role as caustics are formed by the cooperative behavior of families of trajectories and are dissolved by ergodic dynamics.

The rest of this paper is organized as follows: In \cref{sec:HMFModel} we provide examples of violent relaxation in the classical HMF model and formulate a classical hydrodynamic description. In \cref{sec:Mean-field} we describe a theory for the quantum hydrodynamics for the HMF model based on a generalized Gross-Pitaevskii equation (GGPE).  Our numerical solutions of the GGPE are presented in \cref{sec:Numerics} where we explore how quantum effects modify the QSS. In \cref{sec:bicluster} we sketch out the multiscale analysis first used for describing QSS analytically in Ref.\ \cite{Barre2002a},  and show how this is modified by quantum effects. In \cref{sec:dynamicalphasediagram} we explain how we arrived at the non-equilibrium  phase diagram shown in \cref{fig:phase-diagram}, and in \cref{sec:qu-connection} show how the interference patterns decorating the quantum biclusters can be understood using catastrophe theory. 
 In \cref{sec:comm-limits} we argue that the thermodynamic limit $N \rightarrow \infty$ and the classical limit $\hbar \rightarrow 0$ do not commute, and we make our concluding remarks in  \cref{sec:Conclusions}. There are also five appendices covering details omitted in the main text.

\section{Violent relaxation in the classical HMF model \label{sec:HMFModel}}

The HMF model, despite its name, provides an exact description of a many-body system in one dimension. Defined on a ring of radius $R$, it describes $N$ particles interacting via a pair-wise potential varying as $\cos{ (\theta_i-\theta_j) }$, and has the Hamiltonian  
\begin{equation}
  H=\sum_i\frac{L_i^2}{2mR^2} + \frac{\epsilon}{N} \sum_{i<j} \cos(\theta_{i}-\theta_{j}) 
    \label{eq1:intro}
\end{equation}
where each angular momentum $L_i$ and position $\theta_i$ form a canonically conjugate pair $(L_i,\theta_i)$. When two particles sit on top of one another the potential is repulsive (attractive) when $\epsilon>0$ ($\epsilon<0$). The explicit $1/N$ factor in the interaction term, known as the Kac prescription \cite{Kac63}, enforces extensivity of the Hamiltonian \cite{Campa2009}. Experiments with cold atoms trapped in linear optical resonators formed of two mirrors where the atoms interact via the sinusoidal mode of a quasi-resonant optical field \cite{Schutz2014} are described by Hamiltonians closely related to \cref{eq1:intro}. A complementary interpretation of \cref{eq1:intro} is as a kinetic  $XY$ model \footnote{where the kinetic energy of each rotor is included}  with an infinite-range interaction, and therefore, another physical realization is provided by polar molecules in optical lattices \cite{Barnett2006,Yu2009,Carr2009}. 


\begin{figure}[t]
  \begin{subfloat}[Attractive ($\epsilon < 0 $)]
      {  \includegraphics[width=0.455\linewidth]{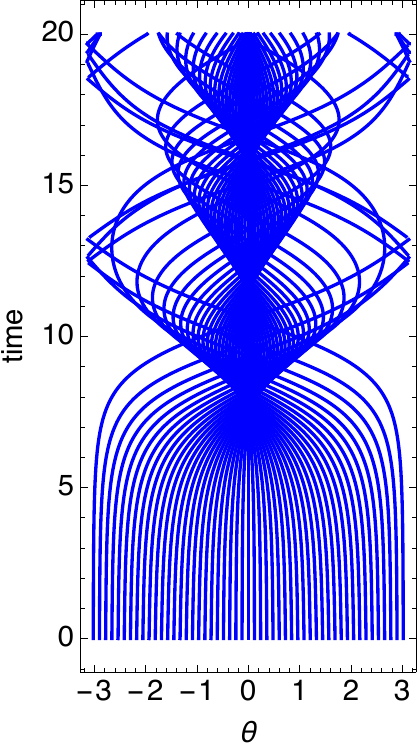}
    \label{fig:newtonian-a}}%
  \end{subfloat}%
  \begin{subfloat}[Repulsive ($\epsilon > 0$)]%
    {  \includegraphics[width=0.49\linewidth]{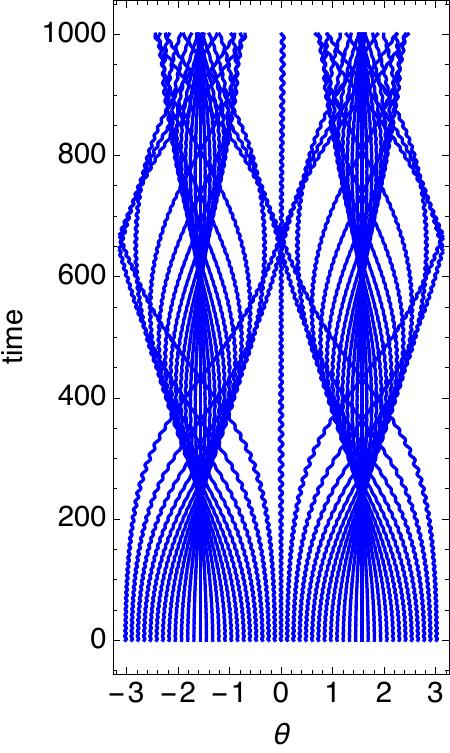}
    \label{fig:newtonian-b}}%
  \end{subfloat}~~~%
  \caption{Long-range interactions tend to amplify initial perturbations. We illustrate this feature here with the Newtonian trajectories of 52 particles obeying the classical HMF model with \protect\subref{fig:newtonian-a} attractive and \protect\subref{fig:newtonian-b} repulsive interactions. At $t=0$ the particles are spaced evenly around the ring with very slightly varying initial velocities $v_i(\theta,0_i)=0.005\cos\theta$. In both cases the LRI cause the particles to cluster and this behaviour repeats such that the envelopes of the trajectories form a quasi-periodic series of cusp-shaped caustics (or ``chevrons''\cite{Barre2002, Barre2002a}). However, there are key differences: Firstly the attractive interactions give rise to a single cluster point (monocluster) around the ring whereas repulsive interactions produce two cluster points (bicluster), and secondly there are very different time scales associated with the two cases with the repulsive case being much slower (note the different scales on the time axes).    
\label{fig:newtonian} }
\end{figure}


There are two distinct types of dynamics in the HMF model which are illustrated in \cref{fig:newtonian}.   The first case has attractive interactions [$\sgn(\epsilon)=-1$] as shown in \cref{fig:newtonian-a}. These lead to a Jeans-like collapse into a monocluster at a single point around the ring which then spreads out and revives periodically, with a timescale that is directly determined by the strength of the inter-particle interaction. The second case has repulsive interactions [$\sgn(\epsilon)=1$] and is shown in \cref{fig:newtonian-b}. One again finds clustering but now at two points on opposite sides of the ring. Furthermore, it arises on vastly longer timescales and corresponds to a QSS.

When the number of particles becomes large a kinetic-theory description in terms of the phase space density $f(\theta, L, t)$ becomes appropriate. The fact that the number of pairwise LRI scales as $\order{N^2}$ whereas collisional terms are $\order{N}$ suggests one can neglect collisions in this regime. In this case $f(\theta, L, t)$ obeys the conservation law \cite{Spohn1992}
\begin{equation}
  \dv{f}{t}=\partial_t f + \dot{\theta}~\partial_\theta f
  + \dot{L}~\partial_L f = 0.
  \label{eq1:Equiv-Classical}
\end{equation}
which is known as the collisionless Boltzmann or Vlasov equation
(see \footnote{The Vlasov equation constitutes a mean-field equation because it deals exclusively with the averaged particle density $f(\theta,L,t)$. Whereas the full Boltzmann equation has knowledge of particle-particle correlations via the collision integral, this is neglected in the Vlasov equation. Finite  $N$ corrections that would appear as collisional terms on the right hand side of \cref{eq1:Equiv-Classical} are suppressed and in the case of the HMF model one can show analytically that these are $o(1/N)$ \cite{Bouchet2005}, while numerical evidence for the HMF model suggests a suppression of $\order{1/N^{1.7}}$ \cite{Yamaguchi2004,Tamarit2005}.} for a discussion of the difference between the Boltzmann and Vlasov equations.) In fact, Hepp and Braun \cite{Braun1977} have rigorously shown that as $N \rightarrow \infty$ the Vlasov equation provides an exact description of the dynamics of an N-body classical system with pairwise LRI.

Biclustering was first identified in numerical simulations \cite{Dauxois00,Firpo02} seeded by a ``water-bag''-shaped initial distribution in phase space: $f(\theta,L,t=0) \propto \Theta( \theta_{0} - \vert \theta \vert) \times \Theta( L_{0} - \vert L \vert)$. In this paper we are interested in the low-temperature regime where the water-bag becomes  thin in the $L$-direction. This both causes the lifetimes of the QSS to diverge in the classical theory \cite{Barre2002a}, and also allows for a natural point of contact between our quantum (low temperature) treatment and the classical dynamics. In the limit $\Delta L \rightarrow 0$ each point in space can be assigned a definite velocity, i.e.\ $f(\theta,L,0)=\rho(\theta)\delta[v(\theta)-L]$, and \Cref{eq1:Equiv-Classical} can be re-expressed in an Euler (i.e.\ hydrodynamic) form. This is what constitutes the zero-temperature approximation.  It is convenient to introduce a new time $\tau=t/\sqrt{mR^2\abs{\epsilon}}$ and velocity $\tv(\theta)=v(\theta)\cdot \sqrt{mR^2\abs{\epsilon}}$, and to normalize the density via $\int \dd\theta \rho = 1$. Written in terms of these quantities, the Euler equations are given by
\begin{subequations}
  \begin{equation}
    \partial_\tau\rho+\partial_\theta(\rho \tv)=0 \label{eq2a:Equiv-Classical}
  \end{equation}
  \begin{equation}
    \partial_\tau \tv +\tv\partial_\theta \tv + 
    \sgn(\epsilon)~\partial_\theta \Phi
    =  0 \label{eq2b:Equiv-Classical}
  \end{equation}
\label{eq2:Equiv-Classical} 
\end{subequations}
and can be interpreted as the equations of motion for a fluid undergoing adiabatic and inviscid flow \cite{Nakayama1997}. Here,
\begin{equation}
  \begin{split}
  \Phi(\theta,\tau)&=\int_{- \pi}^{\pi} \dd \phi \, \rho(\phi,\tau)\cos(\theta-\phi)\\
  &=M(\tau) \cos[\theta-\varphi(\tau)]
  \end{split}
  \label{eq:phi-def-2}
\end{equation}
is the mean-field potential (for the zero-temperature case) found by summing up the long-range interactions amongst all the particles.  It is related to $\dot{L}$ in \cref{eq1:Equiv-Classical} via  the Euler-Lagrange equation $\dot{L}= - \epsilon \partial_{\theta} \Phi$.

The last line of \cref{eq:phi-def-2} represents a remarkable simplification that is proved in \cref{sec:sinusoid}: the mean-field potential always assumes the same cosine functional form specified by just two time-dependent parameters: the depth, or magnetization $M(\tau)$, and a phase $\varphi(\tau)$. Thus, $ \Phi(\theta,t)$ is highly constrained and can only change its depth and position around the ring.

%

\section{Quantum fluid dynamics: Gross-Pitaevskii theory \label{sec:Mean-field} }

 Chavanis \cite{Chavanis2011} has previously investigated the equilibrium properties of the quantum HMF model using the Gross-Pitaevskii theory, finding that when $\epsilon < 0$ quantum effects can stabilize against the Jeans instability. We focus instead on dynamics which take on a heightened importance in the presence of LRI.

 The full quantum description is in terms of a many-body wavefunction $\psi(\theta_1, \theta_2,...,\theta_N,\tau)$, where the set $\{\theta_i\}$ of $N$ independent angles refer to the particle positions.  However, if we consider indistinguishable bosons, in the large-$N$ regime and at very low temperatures, the system can be expected to Bose condense. If all the bosons enter the condensate then  $\psi$ can be written as a product of single-particle wavefunctions [i.e.\ $\psi=\prod_{i=1}^N \varphi(\theta_i)$] which must be found self-consistently due to the effect of interactions. This is the Hartree description and treats the $N$-particle system in terms of a condensate wavefunction, $\psi(\theta_1, \theta_2,...,\theta_N,\tau) \rightarrow \Psi(\theta,\tau)$, which depends on a  single spatial coordinate and obeys a nonlinear wave equation, the Gross-Pitaevskii equation \cite{Pitaevskii2003bose,Pethick2002}. 

Bose condensation therefore naturally leads to a mean field description (equivalent to a hydrodynamic description) and we will assume this situation in our treatment of the quantum problem. In this context it is important to point out that the Mermin-Wagner theorem \cite{Mermin1966,Coleman1973,Mora2001}, which forbids Bose condensation in infinite one-dimensional systems with short range interactions, does not apply here because our system has both finite size and LRI. 
 The mean-field description for a Bose-condensed system is provided by the Gross-Pitaevskii theory which becomes exact in the thermodynamic limit $N \rightarrow \infty$.

\subsection{Generalized Gross-Pitaevskii equation \label{subsec:MFSE}}
 

Consider the Gross-Pitaevskii energy functional 
\begin{equation}
\begin{split}
  E[\Psi,\Psi^{\ast}]=&\frac{N \hbar^{2}}{2m R^2} \int \left|\partial_\theta \Psi\right|^2 
  \mathrm{d}\theta\\
  &+ \frac{ N\epsilon}{2}\int\int \left|\Psi(\theta')\right|^2
  \cos(\theta-\theta')\left|\Psi(\theta)\right|^2\mathrm{d}\theta 
  \mathrm{d}\theta' \, .
  \label{eq1:micro-cann}
\end{split}
\end{equation}
Here the condensate wavefunction is normalized to unity: $\int \mathrm{d}\theta \vert \Psi \vert^2=1$, which ensures that \cref{eq1:micro-cann} is extensive. The equation of motion for $\Psi$ is given by taking the functional derivative $i \hbar \partial \Psi / \partial t = \delta E / \delta \Psi^{\ast}$. One thereby obtains the GGPE for the HMF model \cite{Chavanis2011}
\begin{subequations}
  \begin{equation}
    \iu\tchi \partial_\tau\tPsi=-\frac{\tchi^2}{2}\partial_\theta^2\tPsi + 
    \textrm{sgn}(\epsilon) \Phi(\theta,\tTil)\Psi
    \label{eq2b:MFSE}
  \end{equation}  \begin{equation} \mbox{where} \quad
    \Phi(\theta,\tau)=\int_{-\pi}^{\pi}\left|\tPsi(\phi,\tau)\right|^2
    \cos(\theta-\phi)\mathrm{d}\phi \, 
    \label{eq2c:MFSE}
  \end{equation}
  \label{eq2:MFSE}
\end{subequations}
is the Hartree or mean field potential.  The parameter $\tchi:=\hbar/\sqrt{\abs{\epsilon} mR^2}$ serves as a dimensionless Planck's constant, and we have re-scaled time by introducing $\tau=t/\sqrt{mR^2\abs{\epsilon}}$.  Using the fact that $\left|\tPsi(\phi,\tau)\right|^2$ is the probability density equivalent to the particle density $\trho(\phi,\tau)$ in the zero-temperature classical theory discussed in Section \ref{sec:HMFModel}, the quantum Hartree potential similarly reduces to the 
cosine form given in \cref{eq:phi-def-2}.

For LRI, the Gross-Pitaevskii theory is valid in the high density limit where correlations are weak: for an early discussion in the context of the charged Bose gas see Ref.\ \cite{Foldy1961}.  The validity of the Gross-Pitaevskii treatment with LRI has also been established rigorously for boson stars \cite{Lieb1987} and most recently for dipolar BECs \cite{Triay17}. For a more detailed discussion of the validity of the Gross-Pitaevskii theory for our system see \cref{sec:validity}.

\subsection{Quantum Euler equations \label{subsec:Qu-Euler-Eqn}}
The GGPE can be re-cast in a hydrodynamic form that closely resembles the classical Euler equations. Expressing the generally complex condensate wavefunction as $\tPsi=\sqrt{\trho}\e^{\iu\tS/\tchi}$ we can transform \cref{eq2:MFSE} into two coupled equations describing the evolution of the density $\trho$ and a velocity profile $\tv=\partial_\theta \tS$. The dynamics are then controlled by the quantum Euler, or Madelung, equations \cite{Chavanis2011}
\begin{subequations}
  \begin{equation}
    \partial_\tTil\trho+\partial_\theta(\trho \tv)=0
    \label{eq3a:Qu-Euler-Eqn}
  \end{equation}
  \begin{equation}
   \partial_\tau \tv+ \tv \partial_\theta\tv + 
    \sgn(\epsilon)~\partial_\theta\Phi= -\partial_\theta \left(\frac{\tchi^2}{2}
    \frac{\partial^2_\theta\sqrt{\trho}  }{  \sqrt{\trho} } \right) .
    \label{eq3b:Qu-Euler-Eqn}
  \end{equation}
  \label{eq3:Qu-Euler-Eqn}
\end{subequations}
The expression on the right-hand side is often referred to as the \emph{quantum pressure}; physically it arises from zero-point kinetic energy. Here it is written as the gradient of the so-called quantum potential
\begin{equation}
Q=\frac{\tchi^2}{2}\frac{\partial^2_\theta \sqrt{\rho}}{\sqrt{\rho}}
\label{eq:quantum_potential}
\end{equation}
 and is the only formal mathematical difference between the classical Euler equations, given in \cref{eq2a:Equiv-Classical,eq2b:Equiv-Classical}, and the quantum ones. 


The quantum versus classical aspects of the system can be better appreciated by writing the Gross-Pitaevskii energy functional \cref{eq1:micro-cann} in terms of the hydrodynamic variables $\rho$ and $\tv$ \cite{Chavanis2011}
\begin{equation}
\begin{split}
  E[\Psi,\Psi^{\ast}]&= \frac{1}{2} \langle \tv^2\rangle_\rho +  \frac{1}{2}\langle \Phi \rangle_\rho + \langle Q\rangle_\rho   \\
     &=T_\text{Cl} + U_\text{Cl} +E_\text{Q}
  \label{eq:2-en-func}
\end{split}
\end{equation}
where the averages are taken over the density $\rho(\theta)$. When $\tchi=0$, this agrees with the classical result in the zero temperature approximation. We refer to the contribution of the quantum pressure term  as the quantum energy $E_{Q}$. The remaining terms resemble their classical counterparts: the kinetic energy and the mean-field potential energy are denoted by $T_{\mathrm{Cl}}$ and $U_\text{Cl}$, respectively. 


\subsection{Linear response:  plasma oscillations and Bogoliubov dispersion  \label{sec:Bogoliubov}}
In this work we focus on initial conditions that are close to equilibrium. When violent relaxation occurs it counter-intuitively moves us out of this regime, but linear response still plays a key role at short times.   
Linearizing the hydrodynamic equations \cref{eq3a:Qu-Euler-Eqn,eq3b:Qu-Euler-Eqn} about a spatially homogeneous condensate with density $\rho_0=1/2\pi$ with zero velocity $\tv_0=0$ we obtain
\begin{subequations}
  \begin{equation}
    \partial_\tau \rho_1+ \rho_0\partial_\tau \tv_1=0 
    \label{eq1a:Lin-Qu}
  \end{equation}
  \begin{equation}
   \partial_\tau \tv_1+ 
    \partial_\theta \left(\sgn(\epsilon) \Phi[\rho_1]+
    \frac{\tchi^2}{4\rho_0}\partial_\theta^2\rho_1\right)=0
    \label{eq1b:Lin-Qu}
  \end{equation}
    \label{eq1:Lin-Qu}
\end{subequations}
which describe small excitations about the condensate.
In this linear approximation each Fourier component evolves as an independent oscillator with frequency $\omega_k$ given by the Bogoliubov dispersion relation \cite{Chavanis2011} 
\begin{equation}
  \omega_k^2=\frac{1}{2}\left(\frac{1}{2}\tchi^2k^4+\sgn(\epsilon)~ \delta_{|k|,1} k^2\right).
  \label{eq3:Lin-Qu}
\end{equation}
In the limit $\tchi \rightarrow 0$, and for repulsive interactions ($\epsilon > 1$),  we recover the classical plasma frequency $\omega_\text{pl}=1/\sqrt{2}$ \cite{Barre2002a}, with quantum corrections only appearing at the quadratic level $\omega_{\pm 1}\approx \omega_\text{pl}+\order{\tchi^2}$. The $ \delta_{|k|,1}$  term reflects the fact that with a uniform density only the $k = \pm 1$ modes feel the long-range cosine interaction and are therefore responsible for plasma oscillations. By contrast, the other modes ($k\neq \pm 1$) evolve as free massive particles with frequency $\omega_{k} = (1/2) \tchi \, k^2$, which is a purely quantum effect.  Referring back to the non-equilibrium phase diagram \cref{fig:phase-diagram}, these quantum modes are responsible for the ``free Schr\"{o}dinger'' regime on the right hand side. In the classical case these modes have zero frequency and so take an infinite time to appear.  When the interactions are attractive ($\epsilon <1 $) the frequency is imaginary in the classical limit indicating the Jeans instability, but quantum effects stabilize the system providing $\tchi > \sqrt{2}$ \cite{Chavanis2011}.  

We can estimate the importance of quantum effects by comparing the magnitude of $E_{Q}$ to that of the total classical energy. During classical plasma oscillations the energy alternates between $T_{\mathrm{Cl}}$ and $U_\text{Cl}$ such that these two terms are on average of equal magnitude and we can compare $E_{Q}$ against either of them. We therefore expect classical-like behaviour provided 
\begin{equation}
  \abs{\frac{ E_\text{Q} }{ U_\text{Cl}  }  }\ll 1 \quad \text{(short times)},
  \label{eq:energy-short}
\end{equation}
where we have stressed that the above argument applies on time scales on the order of the inverse Bogoliubov frequency \cref{eq3:Lin-Qu}. In the classical limit this corresponds to the time scale shown in \cref{fig:newtonian-a}, whereas the time scale of the bicluster in \cref{fig:newtonian-b} is much longer, and so we may anticipate this criterion to be insufficient in understanding the role of  quantum effects on the bicluster.

\section{Numerical Results: Violent relaxation in the quantum regime \label{sec:Numerics}}
We now present the results of our numerical simulations of the full GGPE for equivalent initial data to that used in \cref{fig:newtonian}. In the case of the bicluster we have to contend with the very different times scales provided by the fast microscopic plasma frequency and slow the revival time. This makes the computation quite challenging, but from the physical point of view this is why biclusters are examples of QSS and hence relevant to understanding late-time behaviour and possible thermalization.  The fate of these structures in the quantum theory is therefore of great interest. The details of our numerical methods can be found in \cref{sec:Numeric-Method}.



\subsection{Attractive interactions ($\epsilon <0$):  Jeans instability in the quantum regime \label{sec:Numerics-ferro}}
\begin{figure}[t]
  \includegraphics[width=\linewidth]{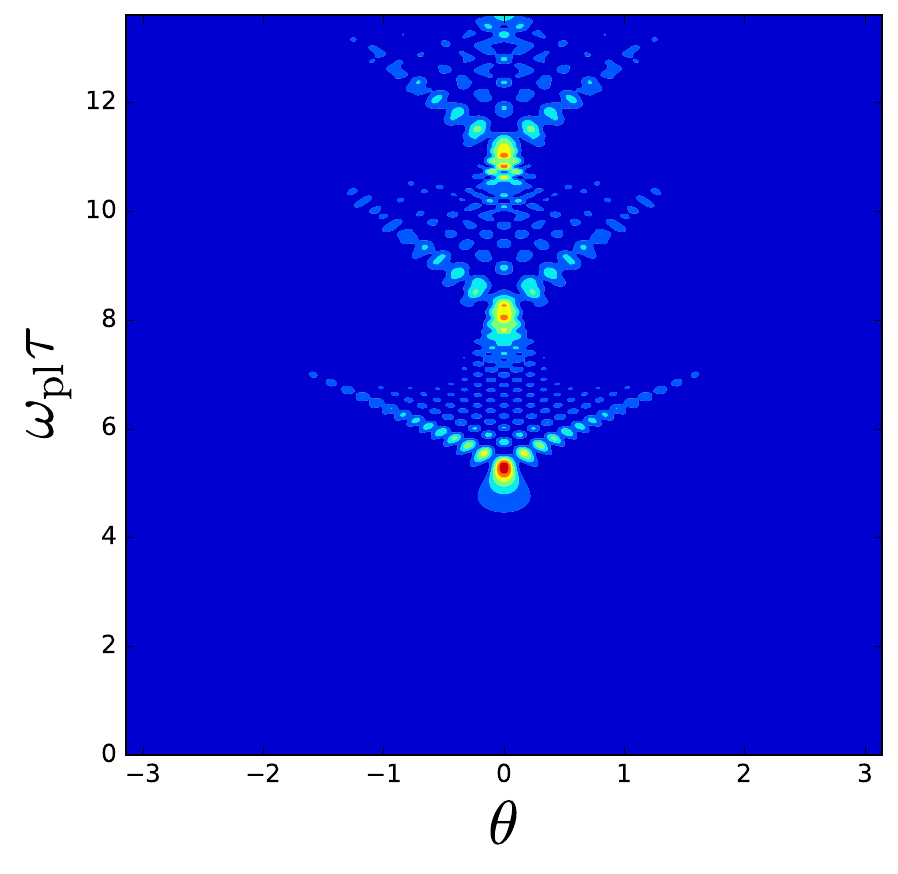}
  \caption{Quantum Jeans instability: dynamics of the density profile  $\rho(\theta,\tau)$ for attractive interactions ($\epsilon <0$) with initial conditions $\rho_0=(1+0.01\cos\theta)^2$ and $\tv_0=0$ with $\tchi=10^{-3}$. Interference tempers the (singular) classical caustic and replaces it with a characteristic interference pattern. The dynamical time scale is set by $\omega_\text{pl}\approx \omega_1 = \text{Im}(\rm{i}/\sqrt{2}) \sqrt{1-\tchi^2/2} $ [\cref{eq3:Lin-Qu}].\label{fig:pearceyDens} }
\end{figure}

In the case of attractive interactions the mean-field potential $\Phi$ favors clustering and this results in the Jeans-like instability. In the classical theory this occurs even for infinitesimal interactions and leads to a cusp caustic with divergent density as shown in \cref{fig:newtonian-a}. However, the Bogoliubov dispersion relation \cref{eq3:Lin-Qu} predicts that quantum zero-point motion stabilizes the system if $\tchi > \sqrt{2}$.  We have confirmed this threshold numerically. An explicit realization of the quantum Jeans instability is presented in \cref{fig:pearceyDens} for $\tchi=10^{-3}$. We see that quantum effects temper the caustic and replace it with an interference pattern such that the density is always finite.

\begin{figure}[t]
  \includegraphics[width=\linewidth]{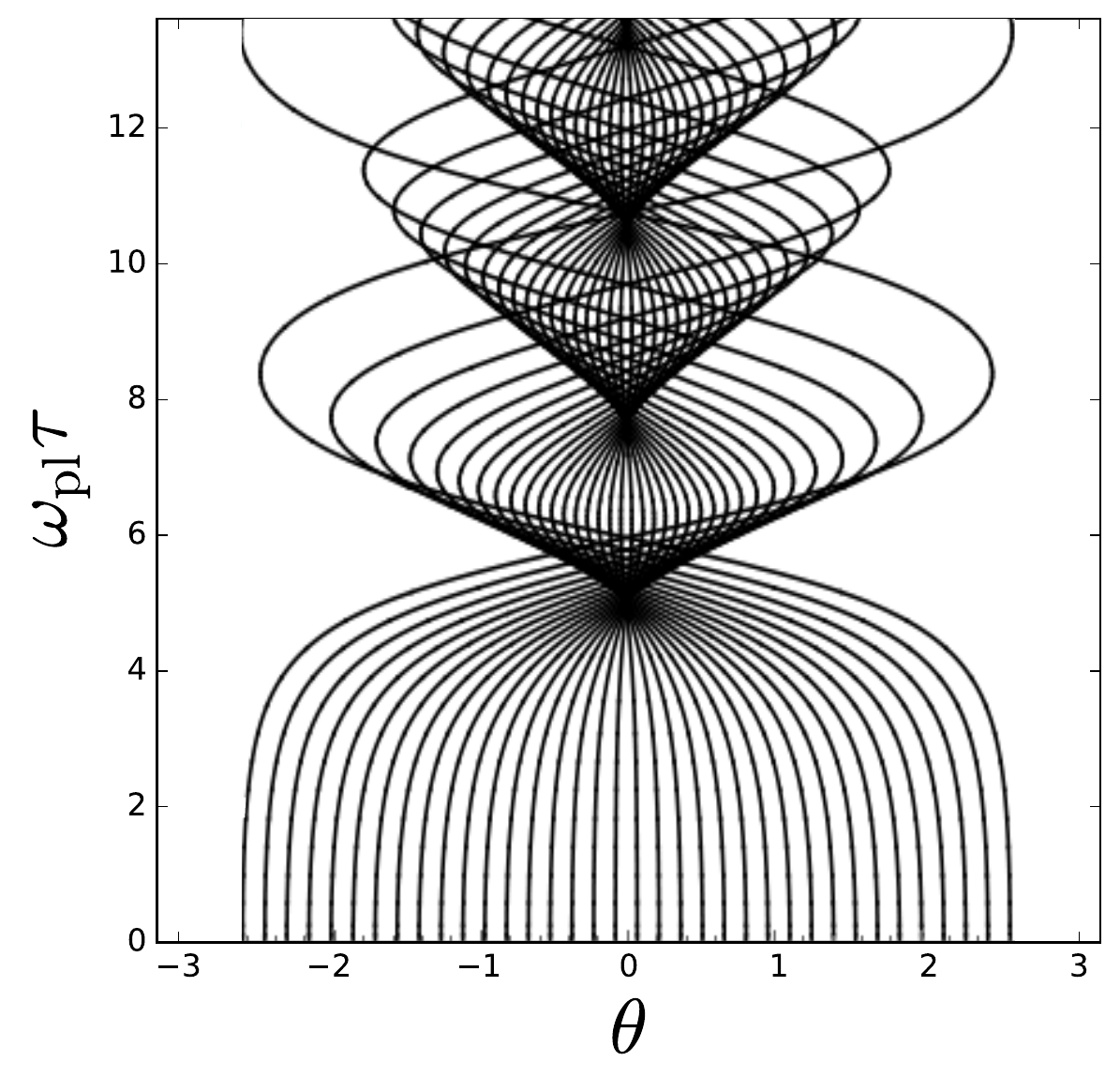}
  \caption{Trajectories of test particles, each computed by solving Newton's equations for a particle moving in a given external potential $V(\theta,\tau)=-M_{\tchi} [\rho(\tau)]\cos\theta$. This is exactly the mean field potential $\Phi(\theta, \tau)$ computed numerically in the quantum dynamics shown in \cref{fig:pearceyDens}, where  $M_{\tchi}[\rho(\tau)]$ is the self-consistent magnetization.    \label{fig:trajectories}  }
\end{figure}



It is important to note that the classical and quantum dynamics only differ qualitatively after the formation of the first cusp, as can be seen by comparing to \cref{fig:trajectories} where we plot the trajectories of non-interacting test particles which simply feel the force generated by the mean-field potential $\Phi(\theta,\tau)$ obtained in making \cref{fig:pearceyDens}  (this is the quantum analogue of the test-particle model discussed in \citer{Levin2014}). This can be easily understood by appealing to energy conservation, and in particular to \cref{eq:2-en-func}. At early times $\left\langle \abs{\partial_\theta\sqrt{\rho}}^2 \right\rangle\lesssim \order{1}$ and consequently $E_Q\lesssim\order{\tchi^2} \ll \abs{U_\text{cl}}$. However the classical dynamics lead to a folding of the phase space distribution (see \cref{fig:cusp-fold}), or equivalently a point-wise divergent density profile, this eventually makes the quantum energy relevant, $E_Q\simeq U_\text{cl}$, at which point interference effects kick in.

\subsection{Repulsive interactions ($\epsilon > 0$): 
bicluster in the quantum regime \label{sec:Numerics-aferro}}

Biclustering is surprising not only because it exists at all (in the presence of repulsive interactions) but also because it occurs at half the wavelength of the mean-field cosine potential. These mysterious features can be explained within the classical theory by using a multi-scale analysis \cite{Barre2002a} that will be discussed in \cref{sec:bicluster}. However, in order to put our quantum results in context it is worth quoting the main result now, namely, that there is an emergent single-particle (i.e.\ non-interacting) description with the effective potential
\begin{equation}
V_\text{eff}=\frac{A^2\omega_\text{pl}^2}{8}\cos 2\theta .
\end{equation}
Apart from the $\cos 2 \theta$ spatial dependence, we note that the depth of $V_\text{eff}$ is proportional to the square of the amplitude of the initial plasma fluctuations, and that this predicts a periodic revival of biclusters with period $T_\text{bc}=\pi/(\sqrt{2} A\omega_\text{pl})$. We shall adopt this as our time scale when plotting the dynamics in the repulsive regime, and note that it is generally much longer than the plasma period used for the attractive case.

\begin{figure}[t]
  \centering
  \begin{subfloat}[\hspace*{3.2em}$ \tchi=5.0\cdot10^{1}$ \newline \hspace*{0.5em}$A\omega_\text{pl}=5.0\cdot10^{-3}$ ]
    {    \includegraphics[clip,width=0.33\linewidth]{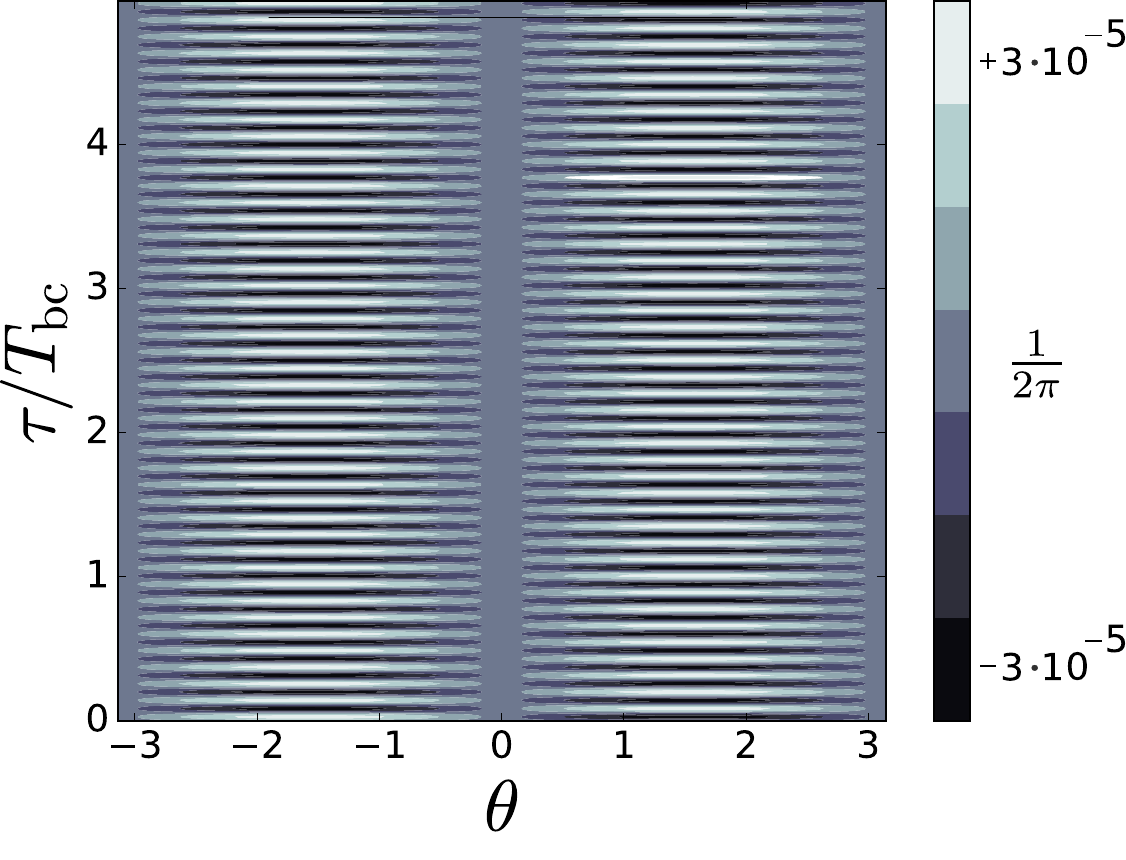}
    \label{fig:schro} }%
  \end{subfloat}%
 \begin{subfloat}[\hspace*{3.5em}$ \tchi=5.0\cdot10^{-1}$ \newline \hspace*{0.5em}$A\omega_\text{pl}=5.0\cdot10^{-3}$ ]
    {    \includegraphics[clip,width=0.33\linewidth]{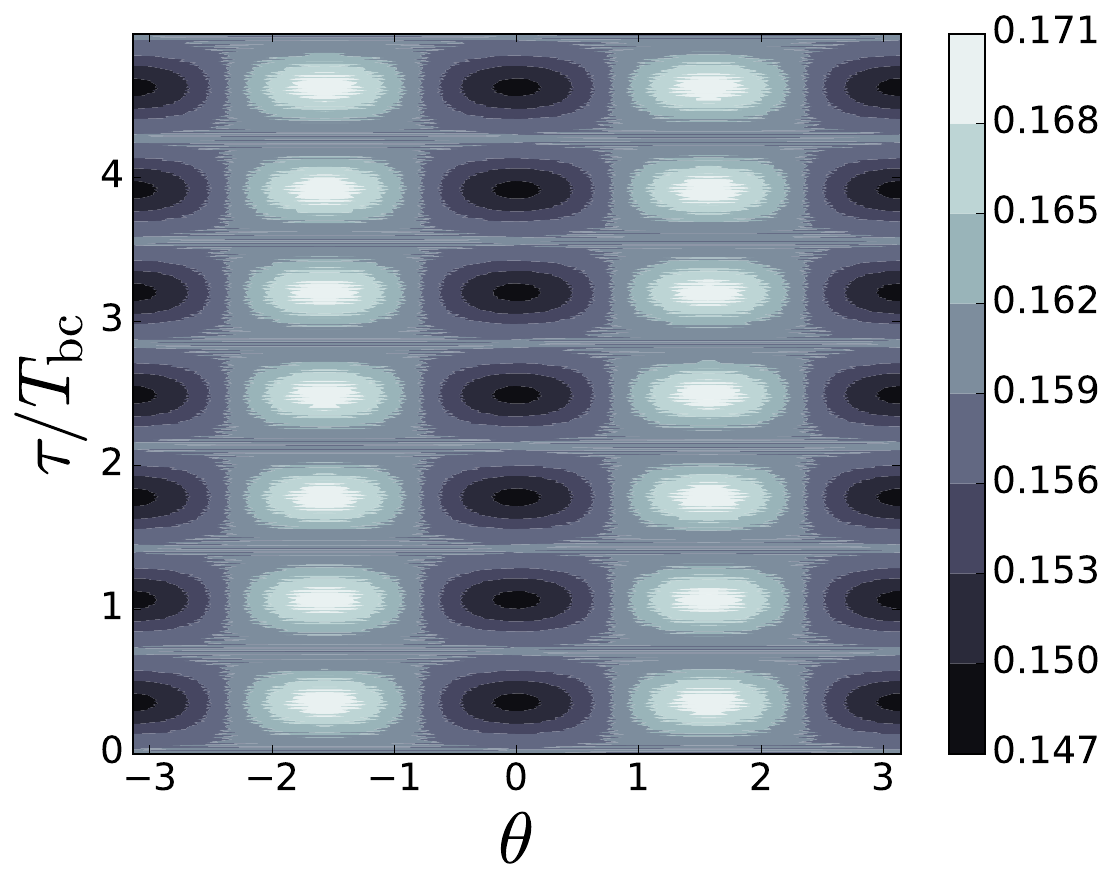}
    \label{fig:plas}}%
 \end{subfloat}%
  \begin{subfloat}[ \hspace*{3.6em}$ \tchi=2.5\cdot10^{-3}$ \newline  \hspace*{0.2em} $A\omega_\text{pl}=5.0\cdot10^{-3}$ ]
    {    \includegraphics[clip,width=0.33\linewidth]{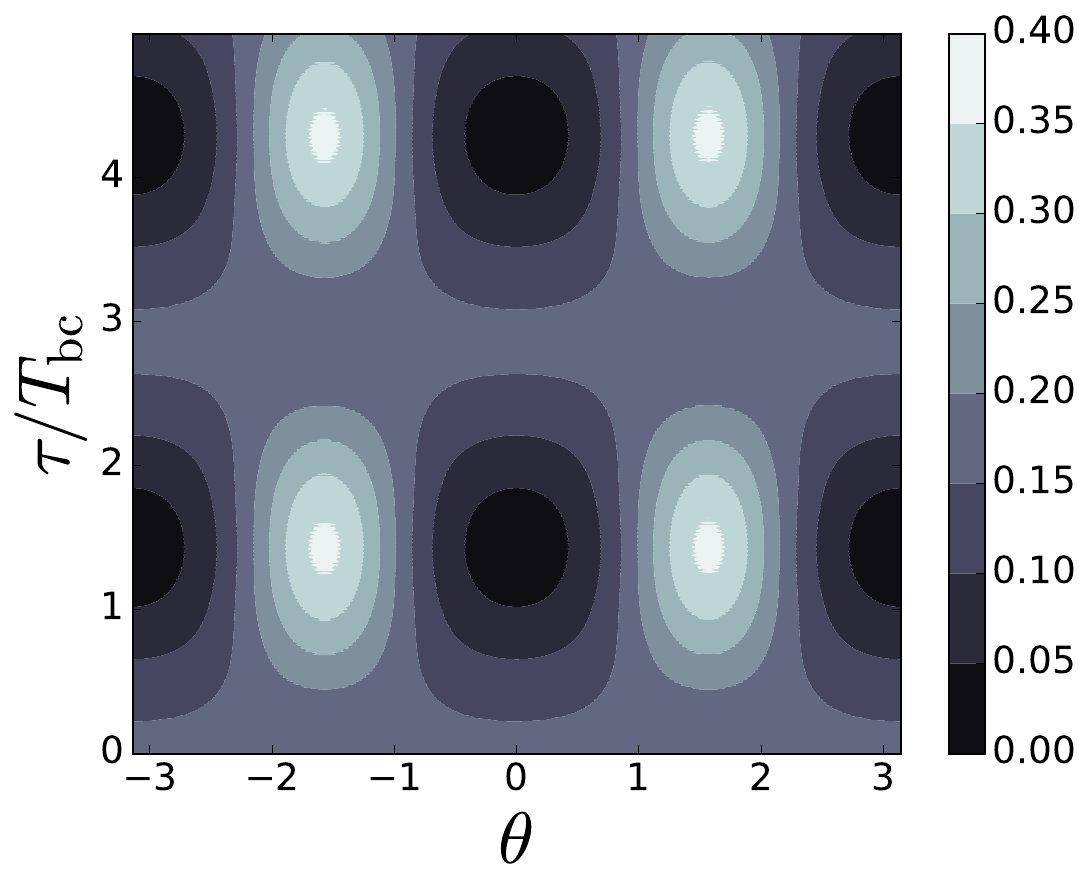}
    \label{fig:bi-a} }%
  \end{subfloat}%

 \begin{subfloat}[\hspace*{3.7em}$ \tchi=2.5\cdot10^{-3}$ \newline \hspace*{0.5em} $A\omega_\text{pl}=1.3\cdot10^{-2}$ ]
    {    \includegraphics[clip,width=0.33\linewidth]{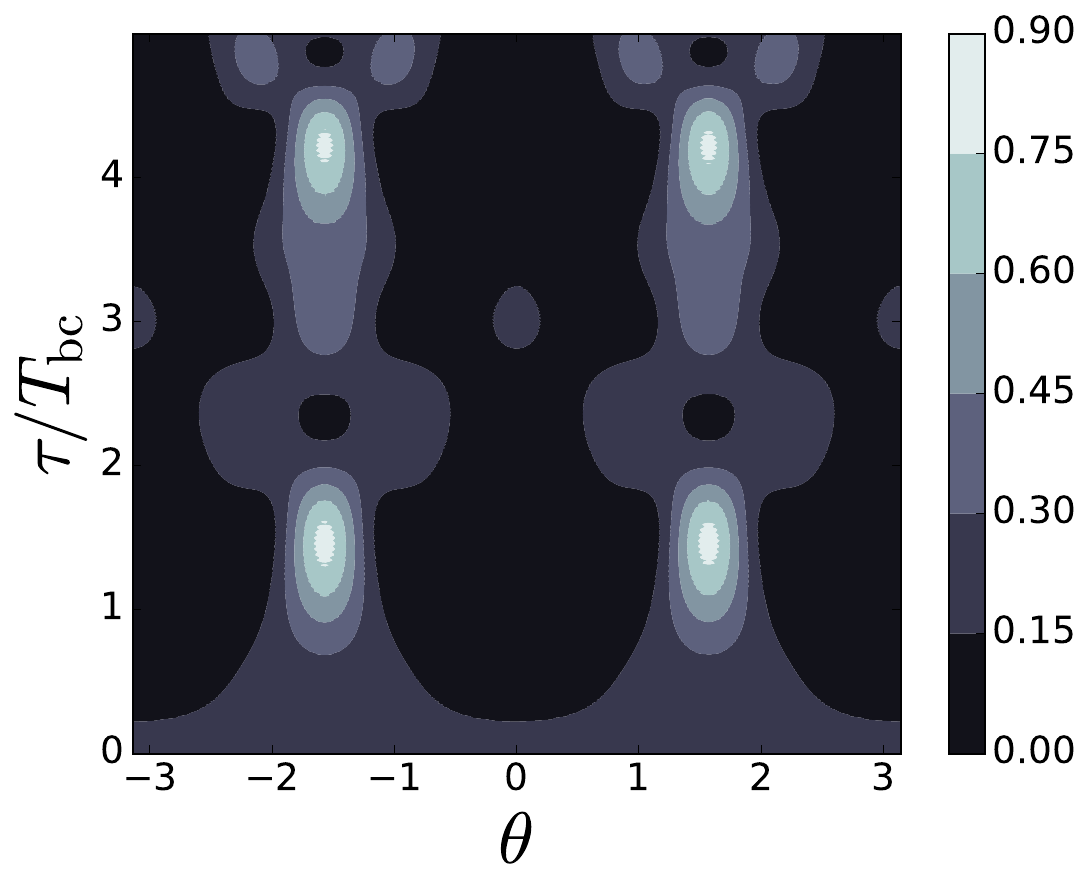}
    \label{fig:bi-b}}%
 \end{subfloat}%
 \begin{subfloat}[\hspace*{3.9em}$ \tchi=2.5\cdot10^{-3}$ \newline \hspace*{0.5em} $A\omega_\text{pl}=2.5\cdot10^{-2}$]
    {    \includegraphics[clip,width=0.33\linewidth]{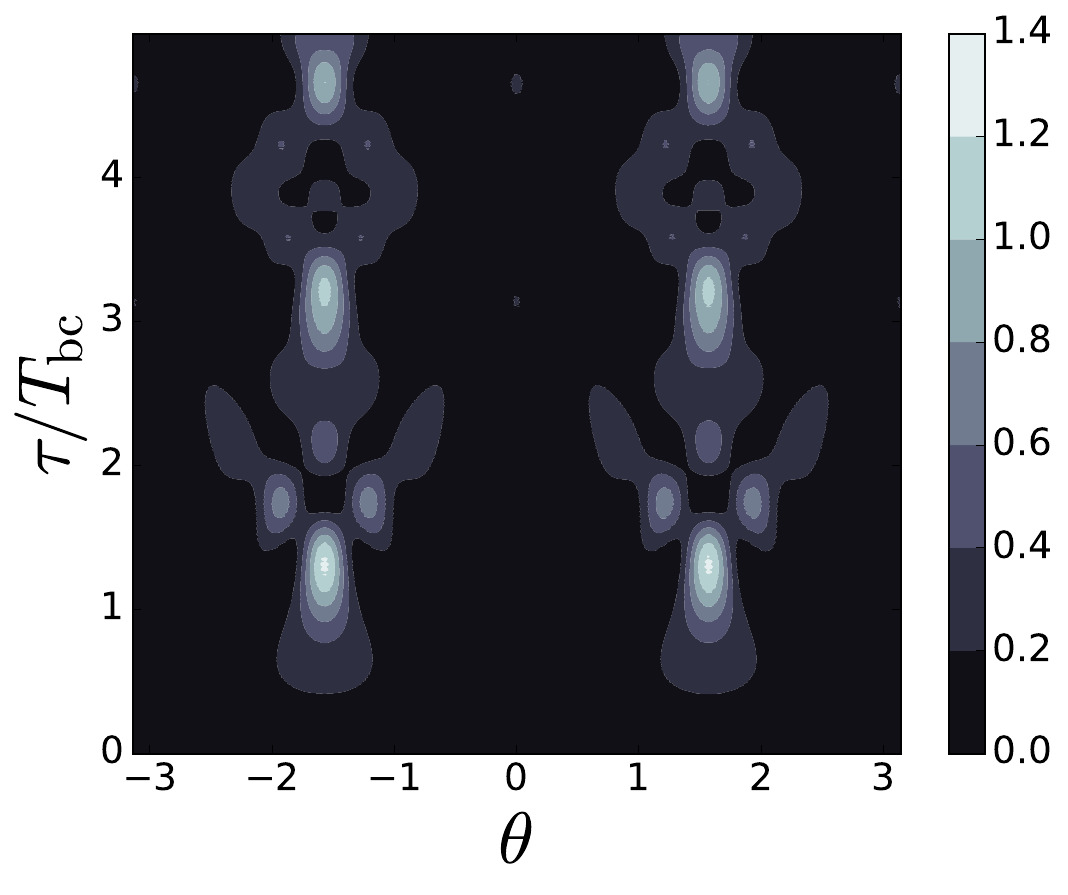}
    \label{fig:bi-c}}%
  \end{subfloat}%
 \begin{subfloat}[\hspace*{3.9em}$ \tchi=2.5\cdot10^{-3}$ \newline \hspace*{0.5em} $A\omega_\text{pl}=5.0\cdot10^{-2}$ ]
    {    \includegraphics[clip,width=0.33\linewidth]{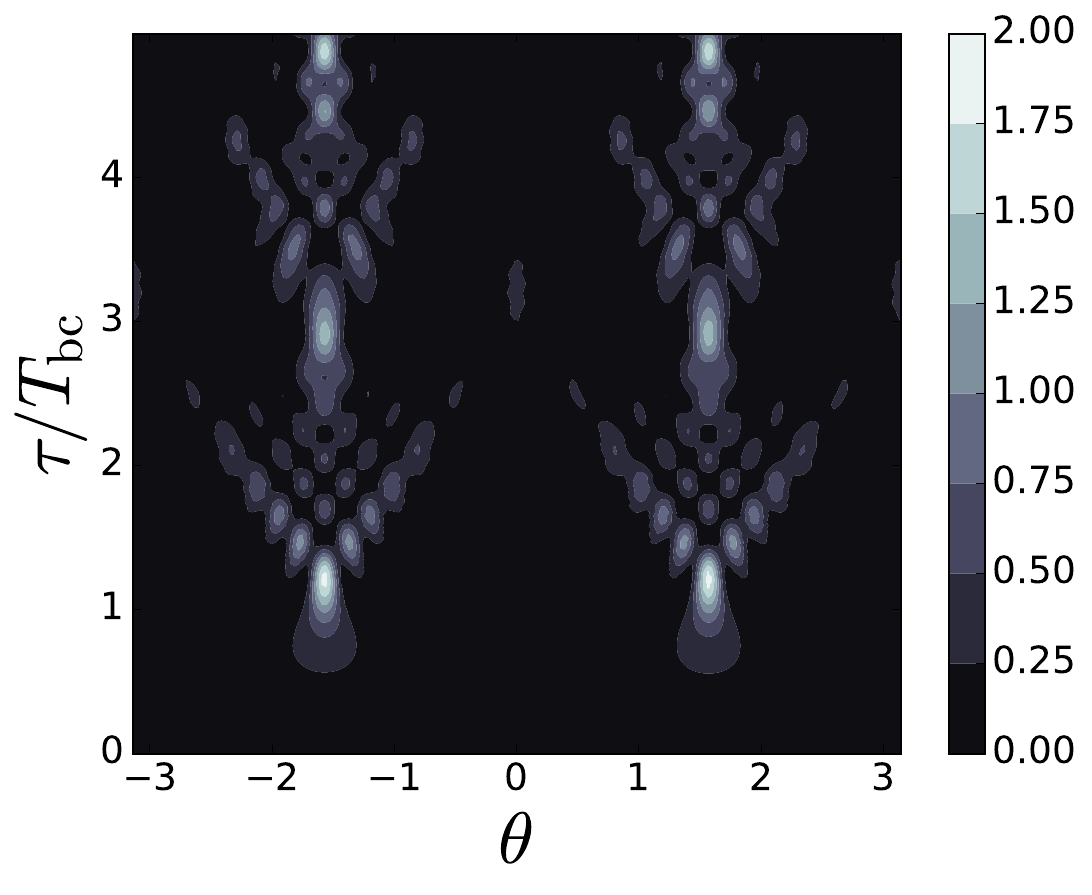}
    \label{fig:bi-d}}%
  \end{subfloat}%
 \caption{Dynamics of the density profile $\rho(\theta, \tau)$ for $\epsilon > 0$ in the $\log\tchi-A\omega_\text{pl}$ plane in correspondence with the different regimes plotted in \cref{fig:phase-diagram}.  The initial conditions in all panels are $\rho_0=1/2\pi$ and $\tv_0=A\omega_\text{pl} \cos\theta$ and time has been rescaled by $T_{\mathrm{bc}}$. For $\tchi\gg 1$ no bicluster develops \protect\subref{fig:schro}. For $A\omega_\text{pl}<\tchi\ll 1$  \protect\subref{fig:plas} classical plasma oscillations occur, and a very weak $\pi$ periodic focusing occurs, but is dispersed by the quantum pressure before the classical focusing time $T_\text{bc}$. In panels \protect\subref{fig:bi-a}-\protect\subref{fig:bi-d} $\tchi$ is held fixed while the plasma amplitude $A\omega_\text{pl}$ is tuned. This has the effect of a deeper effective potential $V_\text{eff}$ [see \cref{eq:qu-multi-scale}], and consequently a more classical-like pattern emerges. This change of behavior is captured by the non-equilibrium phase diagram shown in \cref{fig:phase-diagram}, and is discussed at length in \cref{sec:dynamicalphasediagram}.  \label{fig:barre-chevrons}}   
\end{figure}

In \cref{fig:barre-chevrons} we plot solutions of the GGPE in the repulsive case. In the top row we vary the effective Planck constant $\chi$, and in the bottom row we vary the magnitude of the initial velocity perturbations  $\tv_0(\theta)=A\omega_\text{pl}\cos\theta$.   

\newpage

\onecolumngrid

\twocolumngrid

\noindent All plots have an initially homogeneous density profile $\rho_0=1/2\pi$. The top left hand panel \cref{fig:schro} corresponds to the strongly quantum regime $\tchi \ge 1$ where we see that the clustering has been almost completely eliminated. Moving to the right $\tchi$ is decreased towards the semiclassical regime and biclustering appears, although this by itself does not lead to structures closely resembling the classical case and the temporal period is quite different from the classical bicluster formation time $T_\text{bc}$. In fact, to retrieve something resembling the classical behavior we also need the initial plasma wave's amplitude $A\omega_\text{pl}$ to not be too small as illustrated by the bottom row in \cref{fig:barre-chevrons}. We emphasize that this increase in classical behaviour as  $A\omega_\text{pl}$ is increased  occurs for a fixed value of  $\tchi$.

Our numerical simulations suggest that while the semi-classical condition identified on energetic grounds in  \cref{eq:energy-short}  may be necessary for realizing violent relaxation  in the quantum regime it is certainly not sufficient: one also needs a perturbation away from equilibrium that is large enough to overcome quantum fluctuations.

\section{Quantum pressure and the bicluster \label{sec:bicluster}}

As alluded to above, the emergence of the Jeans-instability for attractive interactions can be accurately diagnosed via a straight-forward linearization of the quantum problem (Bogoliubov theory). Furthermore, the criterion for classical behavior given in \cref{eq:energy-short} is also obeyed in the presence of attractive interactions. By contrast, the repulsive case is more subtle and the bicluster's underlying mechanism is inherently nonlinear. It therefore requires a more sophisticated analysis even in the classical regime. To understand this behaviour we first sketch (details are relegated to \cref{sec:Cl-rev}) the derivation of the effective potential $V_\text{eff}(\theta)$ in the classical limit, which was first presented in \cite{Barre2002,Barre2002a}, and argue that the same procedure can be carried out in the quantum regime provided $\tchi\lesssim 1$. This then provides an effective single-particle picture where semiclassical intuition applies. 

\begin{figure}
  \includegraphics[width=0.8\linewidth]{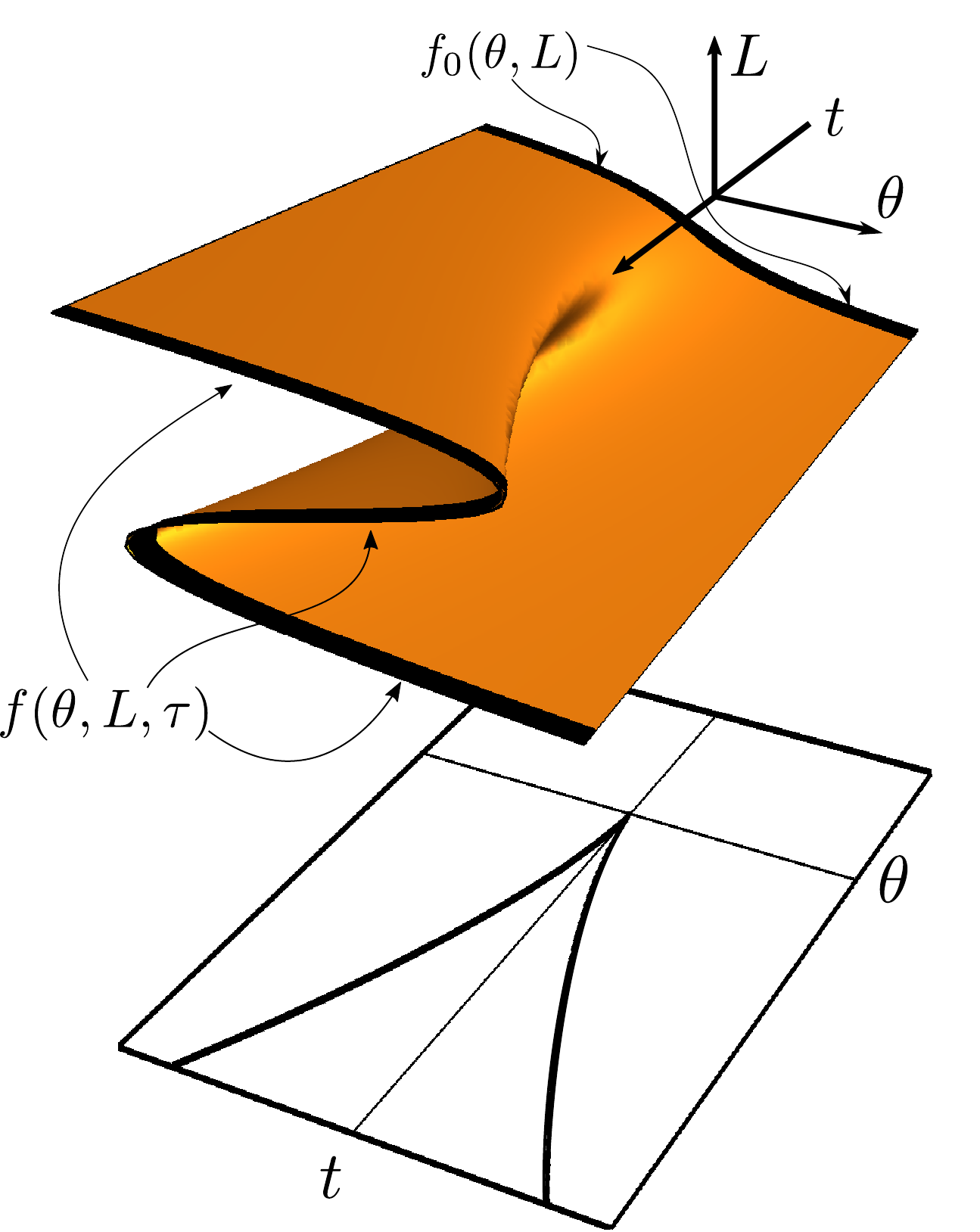}
  \caption{The formation of a cusp catastrophe: illustration of how a line of initial data in phase-space (i.e.\ a vanishingly thin waterbag distribution)  is folded over by the dynamics. Projecting down onto the $\theta-t$ plane produces a cusp shaped envelope on which the density of trajectories diverges. This cusp catastrophe is structurally stable against perturbations and hence occurs generically without the need for special initial conditions. In the case of the bicluster this folding occurs simultaneously at two symmetric points around the ring.  \label{fig:cusp-fold}}
\end{figure}

Starting from the classical Euler equations \cref{eq2a:Equiv-Classical,eq2b:Equiv-Classical}, and performing the analogous linearization procedure to that outlined in \cref{sec:Bogoliubov}, one finds a set of independent Fourier modes all of which have zero frequency, with the exception of the $k=\pm 1$ modes which oscillate with the plasma frequency $\omega_\text{pl}=\sqrt{2}$. This defines a ``fast'' scale, wherein small oscillations of the first Fourier component take place. Anticipating the bicluster that takes place on long time-scales, a slow variable $\mathcal{T}=A \tau$ is introduced which reflects the fact that the time-scale at which nonlinear effects become important is dictated by $A$, the amplitude of the initial plasma wave. Next, the velocity field may be decomposed into a fast part $\tv_1$ evolving under the linear equations, and a slow part, $\tu(\mathcal{T})$ that is influenced by nonlinear effects (the variations of the density can be neglected). A time average then yields 
\begin{equation}
\begin{split}
&\partial_{\tTau}\tu+\tu\partial_\theta\tu=  - \langle \tv_1\partial_\theta \tv_1 \rangle_\tTau :=-\frac{1}{A^2}\partial_\theta V_\text{eff}(\theta) \\
&\mbox{where}  \quad \quad V_\text{eff}(\theta)= \frac{A^2\omega_\text{pl}^2}{8}\cos2\theta.
\label{eq:cl-multi-scale}
\end{split}
\end{equation}
This is a forced Burgers equation governing the flow of the velocity field. The forcing term involves an effective potential $ V_\text{eff}$ with half the wavelength of the mean field potential $\Phi$. Thus, we have obtained the non-trivial result that the slow variables are governed by a different potential than the original. Burgers' equation is well known to give rise to shock waves where the velocity field becomes multivalued and hence the equation breaks down \cite{Burgers1948,MillerBook} and one needs some physical criterion for determining the fate of the system after the shockwave. In our problem these shockwaves are the caustics associated with clustering. As depicted in \cref{fig:cusp-fold}, they occur  when the ``sheet'' of initial data in phase space folds over. When projected onto the $(\theta,\tau)$-plane we obtain cusp-shaped caustics: there are three trajectories passing through each spatial point inside the cusp and one outside. Two of the trajectories coalesce at each point along the cusp edges (known as fold lines) and three coalesce at the cusp tip, which is the most divergent part of the caustic. Quantum mechanically the cusp is therefore associated with three-wave interference giving rise to the characteristic patterns we observe in \cref{fig:pearceyDens}, and in the more semiclassical panels in \cref{fig:barre-chevrons}, which will be discussed  further in \cref{sec:qu-connection}. This highly coherent (non-ergodic) dynamics  is a consequence of the long-range nature of the two-body potential. 

How do these semiclassical arguments fare in the deep quantum regime where quantum zero-point motion can dominate? The essential ingredient in obtaining the forced Burgers equation is the presence of two well separated time-scales. Therefore, provided $\omega_{k=\pm 1} \gg \omega_{k\neq\pm 1}$, the above analysis is valid with the caveat that we must include the effects of the quantum pressure. This condition is naturally satisfied provided $\tchi\ll 1$, as can be clearly seen from \cref{eq3:Lin-Qu}.  Naively, by appending the full quantum pressure to the right-hand side of \cref{eq:cl-multi-scale} we find 
\begin{equation}
  \begin{split}
    \partial_\mathcal{T} \tu + \tu \partial_\theta \tu &= 
    -\frac{1}{A^2}\partial_\theta \left[V_\text{eff}+Q \right]\\
  &= -\frac{1}{A^2}\partial_\theta \left[ \frac{A^2\omega_1^2}{8}\cos2\theta + 
    \frac{\tchi^2}{2}\frac{\partial_\theta^2\sqrt{\rho} }{\sqrt{\rho}}\right],
    \label{eq:qu-multi-scale}
  \end{split}
\end{equation}
while a more sophisticated analysis would subtract off the linearized part of the quantum pressure whose influence is accounted for by the Bogoliubov dispersion relation \cref{eq3:Lin-Qu}. Nevertheless, \cref{eq:qu-multi-scale} correctly predicts the parametric competition between the quantum pressure and the classical effective potential induced by the time-averaged linear plasma oscillations.

We therefore have two distinct semiclassical limits governing the nonlinear dynamics of the HMF model. On short time scales the condition $\tchi\ll 1$ is sufficient to ensure that plasma-like oscillations occur, while the much more stringent condition that $\tchi\ll 2A\omega_\text{pl}$ is required to ensure that the quantum-pressure does not disperse the bicluster. As before, this condition can be re-expressed in terms of energetics where it assumes the form  
\begin{equation}
\begin{split}
  \abs{\frac{ E_\text{Q} }{ U_\text{cl}  }  }&\ll 1 \quad \text{(short times)}  \\
  \abs{\frac{ E_\text{Q} }{ \langle V_\text{eff} \rangle_\rho  }  }&\ll 1 \quad \text{(long times)}.  \\
  \label{eq:energy-long}
\end{split}
\end{equation}
We emphasize that while the above analysis uses a sinusoidal velocity profile as an initial condition, the results are not overly sensitive to this choice. This parallels the classical case, where a thin, but finite, spread in the momentum of an initial water bag distribution still leads to the clustering phenomenon discussed above. Likewise, in the quantum case, deviations from the initial conditions chosen above do not have a dramatic effect on the dynamics.

\section{Non-equilibrium phase diagram \label{sec:dynamicalphasediagram}}

We have already highlighted the fact that systems with LRI take an anomalously long time to come to equilibrium and therefore, rather than equilibrium states, they are  characterized by their QSS.
This motivates the out-of-equilibrium phase diagram presented in \cref{fig:phase-diagram} which we now explain. 

The horizontal axis measures the effective Planck constant $\tchi$ and the vertical axis measures the initial amplitude of the perturbation from equilibrium due to plasma oscillations. This is equivalent to a dependence on the initial energy of the system.  As we are working with a closed system with conserved energy, we may interpret this behaviour in the microcanonical ensemble as a proxy for temperature in the canonical ensemble. We emphasize, however, that the bicluster itself is \emph{not} predicted by a canonical treatment, i.e.\ a system at equilibrium with a heat bath \cite{Dauxois2002}. Rather, this behaviour is inherently non-equilibrium and driven by the long-range interacting nature of the HMF model.

To distinguish the possible regimes  it useful to return to \cref{fig:barre-chevrons}, where results are shown first for $A\omega_\text{pl}$ fixed as $\tchi$ is tuned (\cref{fig:schro,fig:plas,fig:bi-a}), and subsequently for $\tchi$ fixed as $A\omega_\text{pl}$ is tuned (\cref{fig:bi-a,fig:bi-b,fig:bi-c,fig:bi-d}). 
In the first three figures, the most prominent feature is the changing time-scale, which is a consequence of \cref{eq3:Lin-Qu}, and in particular the dependence of $\omega_{k=\pm 1}$ on $\tchi$. Additionally, it is clear that the amplitude of modulations is dramatically different between the three figures, and this is most easily understood on energetic grounds. Initially all three simulations have all of their energy stored as $T_\text{cl}=\frac{1}{4} A^2\omega_\text{pl}^2$. In each case, however, the energetic cost of density modulations is very different. In \cref{fig:schro} the system behaves essentially as a free-Schr\"odinger equation which forms a standing wave, such that $\rho_\text{max} \lesssim A$, while in contrast both \cref{fig:plas,fig:bi-a} are driven at least partially by the interplay between linear plasma oscillations and nonlinear effects as evidenced by the excitation of a $\pi$-periodic density wave. The consequences are markedly different however in that the density modulations in \cref{fig:plas} are a perturbation about a homogeneous background, whereas in \cref{fig:bi-a} they are the $\order{1}$ effect that dominates the density profile, and which signals the onset of nonlinear effects. 

In \cref{fig:bi-a,fig:bi-b,fig:bi-c,fig:bi-d} we can see the emergence of the wave version of the cusp catastrophe as $A\omega_\text{pl}$ is made larger and larger. This enhances nonlinear effects allowing them to dominate the free-Schr\"odinger dispersion of the quantum pressure. Eventually a clear bi-modal cusp-like profile emerges, which signals the validity of the time-averaged treatment, and by association the presence of violent relaxation. 

With these features in hand, we may construct a non-equilibrium phase diagram shown in \cref{fig:phase-diagram}. Note that our initial conditions are limited to the linear regime so that two well-separated time scales exist and we can perform a multi-scale analysis. The crossover between the biclustered regime, and the plasma oscillation regime occurs when $\tchi\approx 2A\omega_\text{pl}$. This is predicted by \cref{eq:qu-multi-scale} and confirmed by the emergence of $\order{1}$ density fluctuations, but the absence of interference effects, in \cref{fig:bi-a}. The crossover between the plasma-dominated and free-Schr\"odinger regimes is found by considering \cref{eq3:Lin-Qu}, and we take $\tchi\approx 1$. Plotting these, as in \cref{fig:phase-diagram} we see that the free-Schr\"odinger regime does not overlap with the bicluster regime for any combination of  $\tchi$ and $A\omega_\text{pl}$. 



\section{Wave catastrophes  \label{sec:qu-connection}}


The cusp-shaped caustics seen in \cref{fig:newtonian,fig:trajectories} result from (imperfect) focusing of classical trajectories, and they are described by Thom's famous catastrophe theory \cite{Thom75,Arnold1975,Poston1978}.  The utility of this theory is that, for each dimension, structurally stable singularities only take on certain universal shapes. Structural stability implies stability against perturbations and hence these objects occur in a wide range of physical phenomena without the need for special symmetry or fine tuning. This universality also extends into the wave/quantum realm where catastrophes give rise to wave patterns known as wave catastrophes or diffraction integrals \cite{Berry1976}. Using a path integral approach provides a rather well defined connection between the classical and quantum dynamics which we now discuss; details can be found in \cref{sec:caustic-analytic}.

Catastrophes are organized into a hierarchy specified by their codimension, with the higher catastrophes containing the lower ones. The simplest is the fold which is generated by the cubic function $S_{f}(C_{1},s) = C_1 s+ s^3$, while the cusp, which is made of two folds, is generated by a quartic function $S_{c}(C_{1},C_{2};s) = C_1 s + C_2 s^2 + s^4$. The parameters $\{C_{1},C_{2}\}$ are known as control parameters: the fold has one, whereas the cusp has two. In the present problem these are the spatial and temporal coordinates, while the state variable $s$ parameterizes paths (e.g.\ the initial angles $\theta_{0}$ around the ring at $\tau=0$, see \cref{fig:trajectories} and also \cref{sec:caustic-analytic}).  In more physical language, a generating function is an action and its saddles give rise to the classical paths via the principle of stationary action $\partial S/\partial s = 0$. Catastrophes are associated with \emph{coalescing saddles}. The fold, being cubic, has two possible stationary points which coalesce on the caustic itself at $C_{1}=0$. The cusp has three possible stationary points: these coalesce in pairs as one crosses either of the two fold lines specified by $C_{1} = \pm \sqrt{8/27} (-C_{2})^{3/2} $, and the most singular point is the tip of the cusp at $C_{1}=C_{2}=0$ where all three stationary points coalesce together.

\begin{figure}[t]
  \begin{subfloat}[Cusp catastrophe]%
    {  \includegraphics[width=0.475\linewidth]{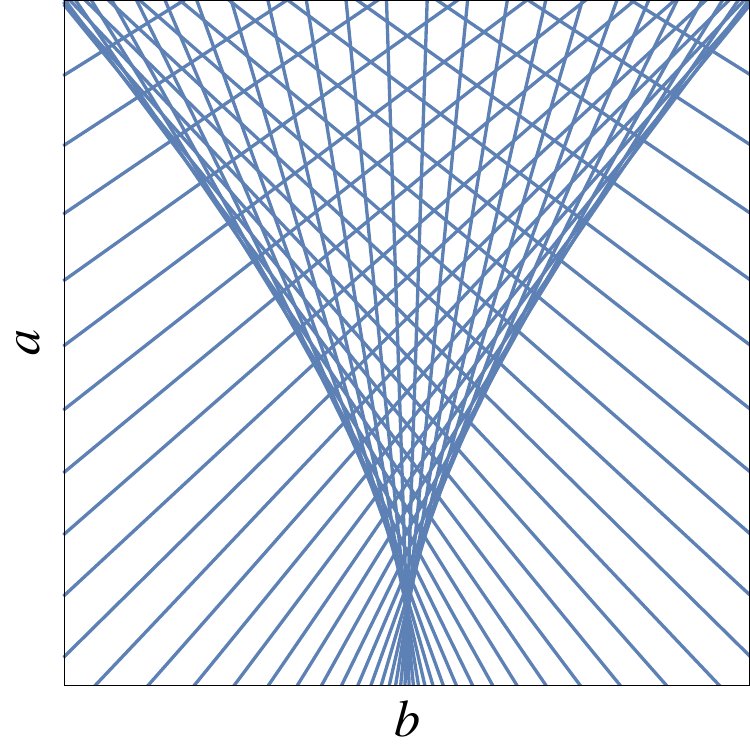} \label{fig:pearcey-1a}}
  \end{subfloat}
  \begin{subfloat}[Pearcey function]
    {  \includegraphics[width=0.475\linewidth]{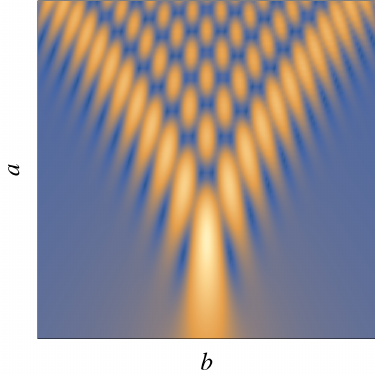}
    \label{fig:pearcey-1b}}
  \end{subfloat}%
  \caption{Classical trajectories in two dimensions will generically form cusp-shaped caustics where the density of trajectories diverges, as shown in \protect\subref{fig:pearcey-1a}. In the wave/quantum theory interference removes the singularity and replaces it with a universal wavefunction, the Pearcey function $\textrm{Pe}(a,b)$, which is valid in the immediate locale of the caustic. In \protect\subref{fig:pearcey-1b} we plot $\abs{\textrm{Pe}(a,b)}$. Note that this  function contains interesting sub-wavelength features such as vortices at its nodes.\label{fig:pearcey}  } 
\end{figure} 

\begin{figure}[ht]
  \includegraphics[width=\linewidth]{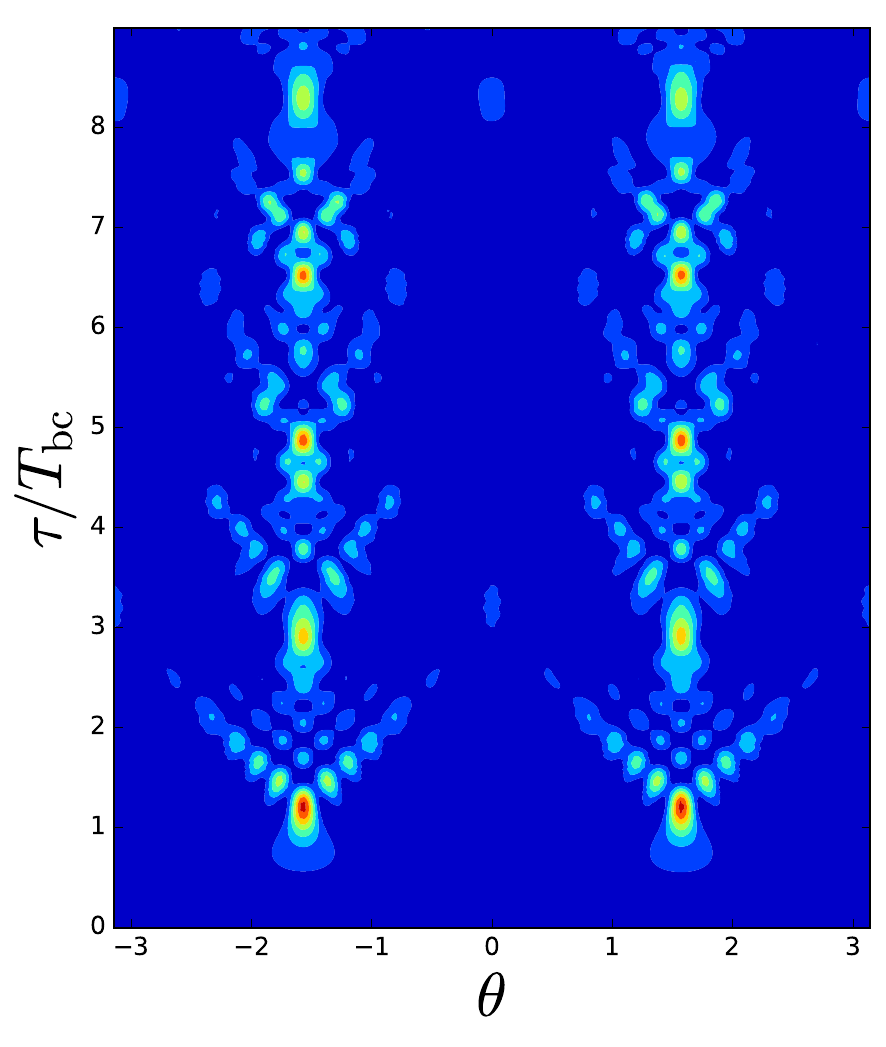}
  \caption{Quantum biclusters: dynamics of the density profile $\rho(\theta,\tau)$ for repulsive interactions ($\epsilon >0$) with initial conditions  $\rho_0=1/2\pi$ and $\tv_0=A\omega_\text{pl} \cos\theta$.  This simulation used $A \omega_\text{pl}=0.01$, $\tchi=0.005$, and included Fourier components up to $k_\text{max}=58$. The time scale is in units of the classical bicluster formation time, $T_\text{bc}=\order{[A \omega_\text{pl}]^{-1}}$ which is two orders of magnitude larger than the inverse plasma frequency $\omega_{\text{pl}}^{-1}$ relevant in the attractive case.
    \label{fig:bi-clust-long}}
\end{figure}

The fact that a catastrophe can be expressed in terms of an action provides a route to quantization motivated by the Feynman path integral prescription. Here one sums over all paths, not just the classical ones, and the amplitude associated with each path is $\exp (\iu S/\hbar)$. In this way one obtains the wave catastrophes \cite{Berry1976}
\begin{equation}
\Psi(\mathbf{C}) = \frac{1}{\sqrt{\hbar}} \int_{-\infty}^{\infty} \e^{\iu S(\mathbf{C}; \mathbf{s})/\hbar } \ \dd\mathbf{s} \ ,
\end{equation}
which are the universal wave functions replacing the divergent classical catastrophes. In the case of the fold the cubic action gives rise to the Airy function $\mathrm{Ai}(x)$ \cite{Berry1980}
\begin{eqnarray}
\Psi_{f}(C_{1}) & = & \frac{1}{\sqrt{\hbar}}  \int_{-\infty}^{\infty} \e^{\iu (C_{1} s+s^3)/\hbar } \ \dd s \nonumber  \\
& = & \frac{2 \pi}{ 3^{1/3} \hbar^{1/6} } \ \mathrm{Ai} \left( \frac{C_{1}}{3^{1/3} \hbar^{2/3}} \right) .
\end{eqnarray}
This implies that as $\hbar$ is varied the overall amplitude of the fold caustic diverges as $\hbar^{-1/6}$ while the fringe spacing vanishes as $\hbar^{2/3}$. The exponent 1/6 is known as the Arnold index and the exponent 2/3 is known as the Berry index.

For the cusp one obtains  
\begin{eqnarray}
\Psi_{c}(C_{1},C_{2}) & = &  \frac{1}{\sqrt{\hbar}}    \int_{-\infty}^{\infty} \mathrm{e}^{\mathrm{i} (C_1 s + C_2 s^2 + s^4)/\hbar } \ \dd s  \\
& = &  \frac{1}{\hbar^{1/4}}  \mathrm{Pe}\left(\frac{C_{2}}{\hbar^{1/2}}, \frac{C_{1}}{\hbar^{3/4}} \right)
 \label{eq:cuspwf}
\end{eqnarray}
where
\begin{eqnarray}
\mathrm{Pe} (a,b)  =   \int_{-\infty}^{\infty} \mathrm{e}^{\mathrm{i} (b t + a t^2 + t^4) } \ \dd t 
\label{eq:pearcey}
\end{eqnarray}
is the Pearcey function  \cite{Pearcey1945} which is a two-dimensional complex-valued function that is tabulated in mathematical handbooks \cite{Olver2010}. We can read off the Arnold index for the cusp as being $1/4$ and the two Berry indices are $1/2$ and $3/4$. The cusp caustic generated by classical paths (obeying $\partial \Phi_{c}/\partial s=0$) is plotted in \cref{fig:pearcey-1a} and the Pearcey function in \cref{fig:pearcey-1b}. 

By comparing \cref{fig:pearcey-1b} with \cref{fig:pearceyDens,fig:barre-chevrons,fig:bi-clust-long} we indeed identify the characteristic Pearcey pattern as must be the case on the grounds of structural stability \cite{Poston1978}. Thus, the new length and time scales introduced by quantum effects in the HMF problem are of universal origin and have non-trivial scaling properties that differ from naive expectations based on the Schr\"odinger equation. Replacing $\hbar$ in the above formulae by $\tchi$ shows that the magnitude of the spatial density modulations scale as $\vert \Psi_{c} \vert^2 \sim \chi^{-1/2} $, while the length and time scales vary as $\sim \tchi^{-3/4}$ and $\sim \tchi^{-1/2}$, respectively. These results hold close to the origin of each cluster; further away non-universal effects creep in (finite system size and interference between Pearcey functions). In particular, in \cref{fig:bi-clust-long} we see well-defined Pearcey functions at the first clustering events but as time progresses the interference patterns become altered, which occurs both because of interference with the tails of the earlier Pearcey functions and also because the underlying classical cusps become narrower with time \cite{Barre2002a}.

\section{Commutativity of the thermodynamic and classical limits \label{sec:comm-limits}}

The non-commutativity of the $N\rightarrow \infty$ and $t\rightarrow \infty$ limits for systems with LRIs is well known \cite{Bouchet2005}, and gives rise to the characteristic feature that if $N \rightarrow \infty$ first a non-equilibrium state will never relax to Maxwell-Boltzmann equilibrium. Furthermore, in single-particle quantum mechanics there is an analogous situation for the $\hbar\rightarrow0$ and $t\rightarrow \infty$ limits, such that completely different results are obtained in the semiclassical and adiabatic limits \cite{Berry84} and also in quantum systems whose classical limit is chaotic  \cite{Delande2001}. We shall now discuss  whether the $N\rightarrow \infty$ and $\hbar\rightarrow 0$ limits commute in order to complete the final link between these three important limits. 

A significant hint comes from comparing the classical and quantum Euler equations---\cref{eq2:Equiv-Classical,eq3:Qu-Euler-Eqn} respectively---and noting that the former is obtained from the full Vlasov equation \cref{eq1:Equiv-Classical} via the zero-temperature approximation. The zero-temperature approximation ignores thermal fluctuations of the momentum  $f(\theta,L)\approx \rho(\theta)\delta(\tv(\theta)-L)$ and hence gives rise to a well-defined velocity profile $v$. The same effect is realized in the quantum case by a different mechanism: BEC gives rise to a well defined phase $\tS(\theta)$ and hence a well defined velocity profile via $\tv=\partial_\theta \tS$. In the limit $\tchi \rightarrow 0$ of \cref{eq3:Qu-Euler-Eqn} we obtain exactly the classical Euler equations \cref{eq2:Equiv-Classical}.

This is interesting because for finite time, the classical equations of motion provide an exact description of a quantum system in the $\hbar\rightarrow0$ limit \cite{Delande2001}, while the Vlasov equation \cref{eq1:Equiv-Classical}, which is a mean-field approximation at finite $N$, provides an exact description of the classical dynamics in the thermodynamic ($N\rightarrow\infty$) limit \cite{Braun1977}. Additionally, motivated by work on boson stars \cite{Lieb1987} Chavanis has argued that \cref{eq2:MFSE} is exact in the thermodynamic limit \cite{Chavanis2011}, presumably when restricted to Bose-condensed initial conditions.

\begin{figure}[!t]
  \includegraphics[width=0.9\linewidth]{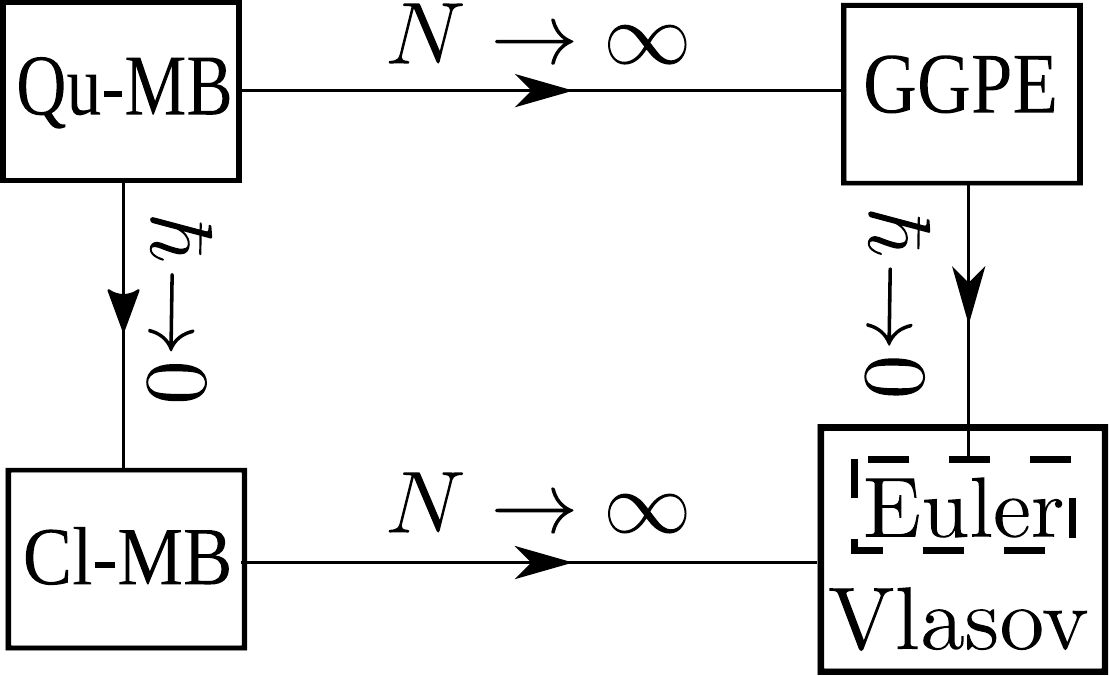}
  \caption{Schematic depiction of non-commutativity of the thermodynamic $N\rightarrow \infty$, and classical $\hbar\rightarrow0$ limits for the finite time dynamics of pure quantum states. The sequence $\hbar\rightarrow 0$ followed by $N\rightarrow \infty$ takes us from the quantum to classical many-body descriptions and then to the Vlasov equation which gives an exact description of the classical dynamics in the thermodynamic limit. Conversely, if the GGPE captures the leading order dynamical behaviour in the thermodynamic limit, then the sequence $N\rightarrow \infty$ followed by $\hbar\rightarrow 0$ leads to the classical Euler equations, which emerge from the Vlasov equation in the zero-temperature approximation. \label{fig:non-commute}}
\end{figure}

Considering a generic quantum state as an initial condition leads to a non-commutativity of limits as illustrated in \cref{fig:non-commute}. In particular, taking the classical limit $\hbar\rightarrow0$, followed by the thermodynamic limit $N\rightarrow \infty$ leads to an exact description in terms of the full Vlasov equation. By contrast, if the GGPE captures the leading order behaviour in the $N\rightarrow \infty$ limit---at least for initially Bose-condensed states---then the Euler equations are obtained; these are only a particular (zero-temperature) limit of the Vlasov equation. This suggests that the recovery of the full Vlasov equation requires features beyond the GGPE, the most obvious of which are phase fluctuations. We note in this context that Chavanis has proposed using the Wigner function to obtain a more complete description of the quantum dynamics \cite{Chavanis2011a}. 


Finally, we note that both $\tchi\rightarrow 0$ and $N\rightarrow \infty$ both cause the density of states of the full quantum many body spectrum to diverge, and so may be expected to yield similar results. The interplay between the three limits $\hbar\rightarrow 0$, $t\rightarrow \infty$, and $N\rightarrow \infty$, for long-range interacting systems is an interesting, and to the authors' knowledge open, problem, and is an obvious avenue of investigation for quantum QMB systems with LRI in general. Its resolution may help shed light on what lessons learned from the study of classical systems with LRI can be carried over into the quantum regime.

\section{Conclusions and Future Prospects \label{sec:Conclusions}}

Motivated by on-going success in the laboratory creating atomic and molecular systems with LRI, we have made a preliminary study of violent relaxation in a quantum system. This non-ergodic dynamical process is a signature of LRI and leads to the formation of slowly evolving patterns with rich and universal structure rather than to the more standard featureless equilibrium state. Although we investigated the dynamics in a specific model, namely the HMF model, it is known to reproduce many of the generic features of many-body systems with LRI.


By choosing initial conditions whose long-time propagation is well understood in the classical limit, we were able to isolate the role played by the quantum pressure in modifying the dynamics. The consequences for attractive interactions ($\epsilon < 0$) are fairly straight forward; whereas the classical self-focusing forms cusp-shaped caustics  in the $\theta-\tau$ plane where the density diverges, these are replaced by smooth but oscillating Pearcey wave catastrophes in the quantum dynamics.  Similar structures have been seen (although they are often not identified as wave catastrophes) in other studies on condensates, both theoretical and experimental, such as BECs hitting obstacles \cite{Carusotto06}, atom optics with BECs \cite{Rooijakkers03,Huckans09,Rosenblum14}, and self-trapping in polariton BECs \cite{Dominici2015}. The underlying connection is non-ergodic dynamics. Indeed, wave catastrophes are expected to be a universal feature of quantum dynamics in mean-field or close to mean-field regimes \cite{Mumford2017}, and more rigorous mathematical analysis of nonlinear Schr\"{o}dinger equations demonstrates this to be true provided the nonlinearity obeys certain constraints \cite{Carles2008,Haberman1985,Hunter1987}. A key implication of the appearance of wave catastrophes in the present problem is the emergence of new spatial and temporal scales that are not present in the classical problem and which scale in a non-trivial way as the effective Planck's constant $\tchi$ is varied.


 The repulsive ($\epsilon > 0$) dynamics also feature cusp caustics although in this case the cusps come in pairs, or biclusters, and the time scale for their formation is generally much longer than in the attractive case. The bicluster has special significance because it is an example of a QSS, i.e.\ a slowly evolving non-equilibrium state that is a paradigm of LRI. As in the attractive case, the biclusters are smoothed by interference and become Pearcey functions. Deeper in the quantum regime zero-point motion becomes dominant and also shifts the time scales for cluster formation and can even stabilize some states against clustering. The bicluster is more sensitive to quantum pressure than the attractive monocluster and the reduced Planck's constant $\tchi$ must be surprisingly small before classical behaviour emerges; specifically the long-time criterion is given by $\tchi\ll A\omega_\text{pl}$, where $A \omega_\text{pl}$ is the amplitude of the velocity fluctuations (``plasma'' waves) in the initial state. At zero temperature one can construct a non-equilibrium phase diagram characterizing the QSS as a function of just $\tchi$ and $A\omega_\text{pl}$ which are the two dimensionless quantities specifying the problem.

In addition to investigating the dynamics, we point out that there is a lack of commutation between the thermodynamic ($N\rightarrow \infty$) and classical ($\tchi \rightarrow 0$) limits. Performing the $\tchi \rightarrow 0$ limit first and then $N\rightarrow \infty$ leads to the Vlasov equation, whereas the opposite order leads to the Euler equations. The latter equations are a special case of the former, corresponding to the zero-temperature limit. We hope that in the future someone will take up the challenge this presents by including non-mean-field quantum states that go beyond the Gross-Pitaevskii theory (at least for finite $N$) and thus examine the implications this has for thermalization.

The HMF model provides a simple arena in which to investigate the essential features of LRI. Atoms trapped in optical cavities come close to realizing the HMF model \cite{Schutz2014} and display a symmetry breaking transition from a homogeneous to an ordered density \cite{Domokos01,Domokos02,Asboth05,Nagy10,Keeling10,Ritsch13,Schutz2014,Schutz2016,Keller2017} that has been observed experimentally \cite{Baumann10} and which is essentially the same phenomenon as clustering.  Although  the atom-cavity system is intrinsically open, and hence includes noise and friction, it would be interesting to see if there are regimes, e.g.\ in very high finesse cavities, where quantum pressure can dominate other sources of noise and stabilize the system against ordering. Another possible realization of the present work is in `closed' XY-type spin models such as those that can be realized with cold Rydberg gases and ensembles of polar molecules \cite{Barnett2006,Yu2009,Carr2009}. Although the interactions in these systems are not infinite-ranged, they can easily extend over the entire sample. Yet another quantum system with LRI, perhaps the most advanced from the experimental point of view \cite{Griesmaier05,Lahaye07,Lahaye08,Koch08,Beaufils08,Lu11,Pasquiou11,Aikawa12,Kadau16,Ferrier-Barbut2016, Chomaz2017,Wenzel2017}, are dipolar BECs. The collapse instability has already been observed in these ``quantum ferrofluids'' \cite{Lahaye08,Koch08}, but it would be interesting to see whether they also display violent relaxation in a geometry where the interactions are predominantly repulsive.

We close by emphasizing that the wave catastrophes (Pearcey functions) studied in this paper are universal features of dynamics. They obey self-similar scaling laws and can therefore be regarded as non-equilibrium generalizations of phase transitions \cite{Mumford2017,Helmrich16}. We hope to expand on this line of inquiry in the future, both for short- and long-range interacting systems.

\section*{Acknowledgements}
We gratefully acknowledge Giovanna Morigi for bringing the HMF model to our attention, and Yan Levin and Julien Barr\'{e} for helping us understand its classical dynamics. Useful insights from both YL and JB were made possible by interactions at ICTP, and both DO and RP thank them for their hospitality. We also thank Robert Dingwall for useful comments on our manuscript, and the OIST school on coherent quantum dynamics which facilitated our interaction. This research was funded by the Natural Sciences and Engineering Research Council of Canada (NSERC) and the Government of Ontario. Support is also acknowledged from the Perimeter Institute for Theoretical Physics. Research at the Perimeter Institute is supported by the Government of Canada through the Department of Innovation, Science and Economic Development and by the Province of Ontario through the Ministry of Research and Innovation.

%

\appendix 

\section{Sinusoidal mean field potential \label{sec:sinusoid}}
In \cref{sec:HMFModel} we claim that $\Phi(\theta,\tau)$ always takes the form of a sinusoid. This result may seem remarkable at first glance but in fact follows from a well known and very simple result. Expanding $\rho(\theta,\tau)=\sum_k \hat{\rho}_k(\tau) \e^{\iu k\theta}$, and writing $\Phi(\theta,\tau)=\textrm{Re}\int_{-\pi}^\pi\rho(\phi,\tau)\e^{\iu (\theta-\phi)}\dd\theta$ one can easily see that $\Phi(\theta,\tau)=2\pi \textrm{Re}\hat{\rho}_1(\tau)\e^{\iu\theta}$. In general $\hat{\rho}_k(\tau)=M(\tau)\e^{\iu\varphi(\tau)}$, where $M(\tau)$ and $\varphi(\tau)$ are two real-valued functions determined by solving for the evolution of the full density profile $\rho(\theta,\tau)$. Simple algebra yields 
\begin{equation}
  \Phi\qty(\theta,\tTil)=M\qty(\tTil)\cos[\theta-\varphi\qty(\tTil)]
  \label{eq4:MFSE-a}
\end{equation}
as claimed in the main text. This derivation applies to both the quantum, and classical Euler equations and can be easily extended to the Vlasov equation by treating a generic phase space density $f(\theta,L,\tau)$ as a linear combination of zero-temperature ones.

\section{Validity of the Gross-Pitaevskii treatment \label{sec:validity}}
It is often stated that Bose condensation in one dimensional systems is forbidden due to phase fluctuations, even at zero temperature \cite{Mora2001}. Formally, Bose condensation implies spontaneous symmetry breaking of the global, and continuous, $U(1)$ symmetry which can be expressed in terms of the field operator as $\hat{\Psi}\rightarrow \e^{\iu\theta}\hat{\Psi}$, and according to the Mermin-Wagner theorem \cite{Mermin1966,Coleman1973},  symmetry breaking is forbidden in one-dimensional systems. However, this theorem  does not apply in finite systems such as a ring of finite radius $R$ where the long wavelength fluctuations which destroy the condensate are cut off by the finite system size. Furthermore, the Mermin-Wagner theorem assumes short-range interactions  and so it does not apply in the presence of LRIs, as evidenced by the fact that the one-dimensional HMF model, quantum or classical, exhibits critical phenomena such as the paramagnetic-ferromagnetic transition \cite{Dauxois2002}.

In typical atomic gases the interatomic potential $V(\mathbf{r}-\mathbf{r}')$ is deep and falls off asymptotically as $r^{-6}$, being of the isotropic van der Waals type, which is considered short-range from a statistical mechanics point of view \cite{Dauxois2002a}. In order to provide a low energy description consistent with the Hartree approximation, $V(\mathbf{r}-\mathbf{r}')$ must be replaced by a pseudo-potential $g \delta(\mathbf{r}-\mathbf{r}')$, where $g = 4 \pi \hbar^2 a_{s}/m$,  $a_{s}$ being the $s$-wave scattering length  \cite{Huang1987}. This reduces the integral in $\Phi$ to a purely local nonlinearity $\propto \vert \Psi \vert^2$ such that the equation of motion for $\Psi$ becomes the usual Gross-Pitaevskii equation \cite{Pitaevskii2003bose,Pethick2002}. By contrast, if the potential is long-ranged and gently varying, like the sinusoidal potential in the HMF model, it is neither necessary nor possible to replace it by a $\delta$-function pseudo-potential and one instead retains the integral in $\Phi$.  The Gross-Pitaevskii equation then takes the form of the integro-differential equation, or GGPE, given in \cref{eq2:MFSE}. This form of GGPE  has previously been successfully employed to treat dipolar Bose gases \cite{Santos2000,Yi2001,ODell2003} and has been rigorously justified in Reference \cite{Triay17}.

 In the presence of long-range interactions it is the high-density regime where Gross-Pitaevskii theory applies. A classic example of this is the charged Bose gas where the criterion for weak correlation is $a_{0} (N/V)^{1/3} \gg 1$, where $a_{0}=4 \pi \epsilon_{0} \hbar^2 / m q^2$ is the Bohr radius associated with the Coulomb interaction between particles of charge $q$ \cite{Foldy1961}. In other words, the interactions are weak if the Bohr radius is large in comparison to the interparticle spacing. A related problem is provided by Boson stars where rigorous analysis has demonstrated that the ground state energy asymptotes to the Hartree value in the thermodynamic limit \cite{Lieb1987}. The high density regime is realized naturally in the $N\rightarrow \infty$ limit, and so it is reasonable to assume that a Gross-Pitaevskii treatment is justified for a system of indistinguishable bosons interacting via LRI.

\section{Numerical method\label{sec:Numeric-Method}}
To solve the evolution of the GGPE we use the momentum space representation of \cref{eq3:MFSE}. 
\begin{subequations}
  \begin{equation}
    \iu\tchi\partial_\tau a_k= \frac{\tchi^2}{2}k^2 a_k + \sgn(\epsilon) \Phi_{kk'} a_{k'}:=\tH_{kk'}a_{k'}
    \label{eq3a:MFSE}
  \end{equation}
  \begin{equation}
    \Phi_{kk'}= \frac{1}{2}
    \left({\bf M}~\delta_{k+1,k'}+ {\bf M}^*~\delta_{k,k'+1} \right)
    \label{eq3b:MFSE}
  \end{equation}
  \begin{equation}
    {\bf M}=\sum_{k\in\mathbb{ Z} } a_k^* a_{k+1}=
    \int_{-\pi}^\pi \left|\Psi\qty(\tTil,\theta)\right|^2\e^{\iu\theta}\mathrm{d}\theta
    \label{eq3c:MFSE}
  \end{equation}
  \label{eq3:MFSE}
\end{subequations}
This approach is advantageous because, unlike a generic two-body potential, the cosine potential only couples adjacent momentum modes. This leads to a tri-diagonal pseudo-Hamiltonian, $\tH_{kk'}(\tau)$ defined in \cref{eq3a:MFSE}, whose time dependence is inherited from the evolution of the order parameter by way of the mean-field potential. The pseudo-Hamiltonian is truncated at $\pm k_\text{max}$ and a second-order implicit integration scheme based on the Dyson series (described below) is used to evolve forward in time.

Given some state $a_k(\tau_n)$ and time $\tau_n$ we first define a time evolution operator $U_{kk'}[\tau_n,\Delta\tau]$ given explicitly by 
\begin{equation}
U(\tau_n,\Delta\tau)=1-i \Delta{\tau} \tH_{kk'}(\tau_n)
\label{eq1:Numeric-Method}
\end{equation}
where $\tH_{kk'}$ is the psuedo-Hamiltonian appearing in \cref{eq3:MFSE} and the
time step is sufficiently small so as not to invalidate Von Neumann error analysis (i.e.\ $\Delta{\tau}< 1/2k_\text{max}^2$). The $0^\text{th}$ order approximation of $a_k(\tau_{n+1})$ is taken to be 
\begin{equation} 
  a^{(0)}_k(\tau_{n+1})=U_{kk'}(\tau_n,\Delta_\tau)a_{k'}(\tau_{n})
  \label{eq2:Numeric-Method}
\end{equation}
Proceeding iteratively, the $m^\text{th}$ approximation is found by solving the equation
\begin{equation}
  U^{(m-1)}_{kk'}\qty(\tau_{n+1},-\tfrac{\Delta\tau}{2})a^{(m)}_{k'}(\tau_{n+1})= U_{kk'}\qty(\tau_{n},\tfrac{\Delta\tau}{2})a_{k'}(\tau_{n})
\label{eq3:Numeric-Method}
\end{equation}
where $U^{(m-1)}$ uses the $(m-1)^{th}$ approximation of $a_k(\tau_{n+1})$. This process is repeated until the overlap between successive states is unity within one part per million. Explicit schemes were also tested, and were found to give monotonically increasing error in the norm of $\Psi$; successful simulation of long-time behaviour considered in this paper (i.e.\ bicluster) requires an implicit scheme. 

Finally we note that to simulate the semi-classical behaviour of the bicluster it is necessary to include an unexpectedly large number of Fourier components. This is related to representing the small amplitude plasma wave $\tv=A\omega_\text{pl}\sin\theta$ in the GGPE form. In particular one must Taylor expand $\exp[i \tS/\tchi]$. Since $\tv=\partial_\theta\tS$ we have $\tS=\order{A\omega_\text{pl}}$. As is discussed in the main text, classical-like behaviour emerges when $A\omega_\text{pl}/\tchi\gg 1$, and to obtain an accurate approximation of the wavefunction's Fourier transform requires that $(A\omega_\text{pl})^{k_\text{max}}/(k_{\text{max}} !)\ll 1$ where $k_\text{max}$ is the largest Fourier component in the simulation. 

\section{Classical description of the formation of the bicluster \label{sec:Cl-rev}}



For repulsive interactions linearized dynamics describing $\{\hat{v}_{\vert k \vert=1}, \hat{\rho}_{\vert k \vert=1}\}$ are oscillatory; following \citer{Barre2002a} we refer to this excitation as a plasma wave. For the initial conditions we choose $\tv_{1}(\theta,0)=A\omega_\text{pl}\cos\theta$ and $\rho(\theta,0)=1/2\pi$, where $A\ll 1$, corresponding to a uniform density plus a small position dependent velocity modulation $\tv_1(\theta)=A\omega_\text{pl}\cos\theta$. In direct analogy with \cref{eq3:Lin-Qu}, the solutions to the linearized equations of motion with these initial conditions are \cite{Barre2002a}
\begin{equation}
  \rho_1=-\frac{A}{2\pi} \sin(\omega_\text{pl}\tau)\sin\theta ~,~ \tv_1=A \, \omega_\text{pl} \cos(\omega_\text{pl}\tau)\cos\theta,
  \label{eq:linsolns}
\end{equation} 
where $\omega_\text{pl}:=\omega_{1}=1/\sqrt{2}$ denotes the plasma wave's frequency. As will be shown below, unlike for attractive interactions, the formation of the bicluster is not driven directly by the rapidly oscillating mean-field potential, but rather by a slowly evolving effective potential $V_\text{eff}$ induced by time-averaged linear plasma oscillations. 

As we first saw in Fig.\ \ref{fig:newtonian},  the bicluster forms very slowly in comparison to the plasma period and this suggests a multiple scales analysis \cite{Barre2002a}. Consequently, we consider $\tv(\theta,\tau)=\tv_1(\theta,\tau)+A\tu(\theta,\mathcal{T})$ where $\mathcal{T}=A\tau$ is $\order{1}$ when the fast time $\tau$ is  $\order{1/A\omega_\text{pl}}$; note this hierarchy is only present for $A\ll 1$. Inserting this expression into \cref{eq2b:Equiv-Classical}, and averaging over many plasma oscillations leads us to
\begin{widetext}
\begin{equation}
  \begin{split}
    A^2\partial_{\tTau} \tu
    -\underbrace{\frac{\omega_\text{pl}^2 A^2}{2}\sin(2\theta)\
    \cos^2(\omega_\text{pl}\tTil)}_{\tv_1\partial_\theta \tv_1}+  A^2 \tu\partial_\theta \tu
    & 
    +\omega_\text{pl}^2 A^2
    \left[\cos\theta \partial_\theta \tu- \tu\sin\theta\right]
    \cos(\omega_\text{pl}\tTil)
    = \underbrace{\int_{-\pi}^\pi \rho(\theta',\tTil) \sin(\theta-\theta')\dd\theta'-\pdv{\tv_{1}}{\tTil}}_{\mathcal{L}}\\
  \text{time averaged} \implies \partial_{\tTau}\tu+\tu\partial_\theta\tu=  - \langle \tv_1\partial_\theta & \tv_1 \rangle_\tTau :=-\frac{1}{A^2}\partial_\theta V_\text{eff}(\theta) \quad  \quad \mbox{where}  \quad \quad V_\text{eff}(\theta)= \qty[\frac{A^2\omega_\text{pl}^2}{8}\cos2\theta].
  \end{split}
  \label{eq4:Lin-Qu}
\end{equation}
\end{widetext}
In going from the first line to second line we have used the fact that the convolution on the right hand side depends only on the first Fourier component of the density $\hat{\rho}_{k=\pm1}$ and consequently the quantity $\mathcal{L}$ vanishes at all times \footnote{The much weaker criterion of (nearly) vanishing on average \unexpanded{$\langle \mathcal{L} \rangle_\tTau$} is sufficient for the associated multi-scale analysis to remain valid.} because it satisfies the linearized equations of motion. This gives a remarkably simple result \cite{Barre2002,Barre2002a}: the slow velocity field $\tu(\theta,\mathcal{T})$ obeys an Euler equation driven by $V_\text{eff}(\theta)$ [compare with Eq.\ (\ref{eq2b:Equiv-Classical}), the original Euler equation obeyed by the full velocity field]. $V_\text{eff}$ is derived from the square of the plasma wave and hence corresponds to a potential with two minima around the ring, giving rise to two symmetric clustering points.  
Furthermore, there is negligible back-action on the plasma wave by the slow dynamics:  $V_\text{eff}$ is invariant under $\theta\rightarrow \theta+\pi$, and consequently only influences $\hat{\rho}_k$ and $\hat{v}_k$ for $k$ even. We therefore expect the linear dynamics of the first Fourier component to continue to be a good approximation even at late times.

The first appearance of the bicluster state can be estimated by considering \cref{eq4:Lin-Qu}, whose characteristics are Jacobi elliptic functions. Near a minimum of $V_\text{eff}$ these are approximately cosine functions, and we can approximate $V_\text{eff}\approx \frac{A^2 \omega_\text{pl}^2}{4}(\theta-\theta_{\mathrm{min}})^2$. This identifies the frequency of the bicluster oscillation in the original fast time coordinate  as $\omega_{\mathrm{eff}}=A\omega_\text{pl}/\sqrt{2}$. Consequently, the first bicluster will appear at one quarter the oscillator period $T_\text{bc}= \pi/2\omega_\text{eff}=\pi/\sqrt{2}A\omega_{\mathrm{pl}}$ \cite{Barre2002a}.

\section{Universality in self focusing \label{sec:caustic-analytic}}
Pearcey functions can be expected as a generic consequence of self focusing in coherent quantum systems. The reasons behind this are most easily understood in the case of linear wave equations, however the same underlying ideas generalize for sufficiently weak non-linear effects. Importantly, the HMF model's non-linearity acts essentially as a linear, albeit time-dependent, background potential, which is relatively insensitive to the local structure of the wavefunction, and therefore its dynamics are well modelled by linear theory.

First, let us consider the one-dimensional linear  Schr\"odinger equation
\begin{equation}
 \iu\hbar \partial_t \psi = -\frac{\hbar^2}{2} \partial_x^2 \psi + V(x) \psi,
\end{equation}
with initial data $\psi(x,0)= \psi_0(x)=A(x)\exp[\iu \varphi(x)/\hbar]$. The solutions of this equation may be expressed via the integral equation 
\begin{equation}
\begin{split}
\psi(x,t) &= \int \dd x' K(x';x,t)\psi_0(x')\\
&= f(t) \int \dd x' A(x')\e^{\iu[S_\text{cl}(x';x,t)+\phi(x')]/\hbar}
\end{split}
\label{eq:Schro-Prop}
\end{equation}
where we have used the fact that the propagator may be expressed in terms of the classical action via  $K(x';x,t)=f(t)\exp[\iu S_\text{cl}(x';x,t)/\hbar]$ \cite{Sakurai1994}. In the limit that $\hbar\rightarrow 0$ the above integral is dominated by the stationary points (with respect to $x'$) of what we will now refer to as the generating function $\Phi=S_\text{cl}+\phi$. 

The integral in \cref{eq:Schro-Prop} is over all paths labelled by their initial positions $x'$ (see \cref{fig:trajectories} for an illustration of the paths.) 
We interpret $x$ and $t$ as \emph{control parameters} of the function $\Phi(x'; x,t)$, and for a given choice of $x$ and $t$ we can generically expect the stationary points of $\Phi$ to be locally quadratic in $x'$. If, however, we consider \emph{all} values of $x$ and $t$  then it is generic that pairs of stationary points will coalesce leaving the function $\Phi$ looking locally cubic (hereafter referred to as a fold), and furthermore these cubic points may coalesce to give locally quartic behaviour $\Phi\propto (x'-y')^4$ (hereafter referred to as a cusp). Remarkably, it is \emph{not} generic that these quartic points may coalesce when one varies two control parameters (in this case $x$ and $t$) \cite{Poston1978}, and therefore these three possibilities (saddle, fold, cusp) are exhaustive in a two dimensional control space. 

The coalescing of saddles is in direct correspondence with self-focusing behaviour in the classical dynamics. This can be understood as follows: a saddle represents a classical trajectory, and a coalescing of saddles represents the focusing of two trajectories. Catastrophe theory guarantees us that when focusing occurs in a two dimensional control space, we need only consider folds and cusps. In an associated wave theory such as the HMF model's GGPE this structure is inherited via an Airy and Pearcey function structure, respectively. 

This can be understood by taking \cref{eq:Schro-Prop} and transforming the control variables $(x,t)\rightarrow (C_1,C_2)$ and introducing $s= (\kappa/\hbar)^{1/4}(x'-y') $ (where $\kappa=\Phi^{(4)}\rvert_{y'}/4!$) such that $\Phi(s; y,\tau)$ assumes its \emph{normal form} \footnote{This is the analog of the method of steepest descent for saddles, where the normal form is a $\Phi(s)\sim \hbar[\frac{1}{2} s^2 + \order{s^3}]$ and results in a Gaussian integral}.  For the cusp catastrophe this is $\Phi(s; C_1, C_2)=\hbar(s^4 + C_2 s^2 + C_1 s)$ and this form is valid when the control parameters $(C_1,C_2)$ [or equivalently $(x,t)$] are close to the cusp point or fold line. One can then expand $A(x')$ about $s=0$ and assuming that $A(x')= A_0 + \order{s}$, and likewise that $f(t)\approx f_0$ in the vicinity of the cusp this leads to the local form of the wavefunction \cite{Olver2010}
\begin{equation}
\begin{split}
 \psi(C_1,C_2) &\sim f_0 A_0 \qty(\frac{\kappa}{\hbar})^{1/4} \int \dd s \exp[\iu (s^4 + C_2 s^2 + C_1 s)] \\
   &= f_0 A_0 \qty(\frac{\kappa}{\hbar})^{1/4}  \textrm{Pe}(C_2,C_1).
\end{split}
\end{equation}
Similar considerations give Airy functions in coordinates perpendicular to the fold lines. 

Importantly  wave catastrophes are \emph{robust} against both perturbations in the initial data, and the precise details of the wave equation. This is the statement of structural stability upon which catastrophe theory is built. This provides a justification for identifying the structures in \cref{fig:pearceyDens,fig:bi-clust-long,fig:bi-c,fig:bi-d} as Pearcey functions, and this claim is further vindicated by the fact that in the classical theory these same structures are well understood to result from the pile-up of trajectories. 

Finally we comment on the robustness of these caustic structures even in the presence of stronger local non-linearities such as the in GPE. One might expect that the universality discussed above would be destroyed by a term such as $\abs{\psi}^2\psi$ since $\abs{\psi}^2\sim \mathcal{O}(1/\sqrt{\hbar})$ near a cusp point.  Surprisingly, however, rigorous mathematical studies have demonstrated that even in this case the linear theory presented above is trustworthy in the vicinity of the caustic for surprisingly large non-linearities. This is true both for Airy-function behaviour near a fold catastrophe \cite{Hunter1987,Carles2008}, and Pearcey function behaviour near a cusp catastrophe \cite{Haberman1985}. The fact that these structures are robust, even against a local non-linearity suggests that linear analysis should certainly hold for the much milder non-local GGPE employed in this work.


%

\end{document}